\documentclass[12pt,a4paper]{article}

\usepackage{bm}%
\usepackage{graphicx}%
\usepackage{multirow}%
\usepackage{amsmath,amssymb,amsfonts}%
\usepackage{amsthm}%
\usepackage{mathrsfs}%
\usepackage[title]{appendix}%
\usepackage{xcolor}%
\usepackage{textcomp}%
\usepackage{manyfoot}%
\usepackage{booktabs}%
\usepackage{algorithm}%
\usepackage{algorithmicx}%
\usepackage{algpseudocode}%
\usepackage{listings}%

\usepackage{natbib}
\usepackage{mathtools}
\usepackage{enumerate}

\theoremstyle{definition}
\newtheorem{example}{Example}

\newcommand{\SFSblue}[1]{{\color{black} #1}} 
\newcommand{\Gblue}{\color{black}}

	\newcommand{\bmWi}{\mathbf{A}}

\newcommand{\by}{\mathbf{y}}

\newcommand{\bb}{\mathbf{b}}
\newcommand{\bB}{\mathbf{B}}

\newcommand{\bS}{\mathbf{S}}
\newcommand{\bR}{\mathbf{R}}
\newcommand{\bI}{\mathbf{I}}

\newcommand{\btheta}{\boldsymbol{\theta}}
\newcommand{\bTheta}{\boldsymbol{\Theta}}
\newcommand{\bvartheta}{\boldsymbol{\vartheta}}
\newcommand{\boldeta}{\boldsymbol{\eta}}
\newcommand{\bmu}{\boldsymbol{\mu}}

\newcommand{\bSigma}{\boldsymbol{\Sigma}}

\newcommand{\bxi}{\boldsymbol{\xi}}
\newcommand{\bpi}{\boldsymbol{\pi}}

\newcommand{\bzeta}{\boldsymbol{\zeta}}
\newcommand{\betad}{d}
\newcommand{\cN}{\mathcal{N}}
\newcommand{\cG}{\mathcal{G}}

\newcommand{\ym}{{\mathbf y}} 

\newcommand{\Prob}[1]{\mbox{\rm Pr}\{#1\}}

\newcommand{\PPR}{PPR} 
\newcommand{\trans}[1]{#1 ^{\top}}
\newcommand{\knot}{k'}
\newcommand{\funcTh}[1]{\varphi(#1)}

\begin{document}
	
\title{CliPS -- How to identify cluster distributions in Bayesian mixture models}

\author{Gertraud Malsiner-Walli \thanks{WU Vienna University of Economics and Business}, 
	Sylvia Fr\"uhwirth-Schnatter
	\thanks{WU Vienna University of Economics and Business}\\
	and  Bettina Gr\"un \thanks{WU Vienna University of Economics and Business}
	\vspace{0.5cm}}

\date{}

\maketitle	

%
%

{\small {We propose the CliPS procedure when fitting Bayesian mixture
		models in the context of model-based clustering to identify the
		cluster distributions while simultaneously assessing the suitability of a cluster
		solution and validating the cluster structure.  The procedure relies
		on the point process representation of a mixture model and is based
		on the assumption that a suitable cluster solution requires the
		clusters to be distinguishable with respect to a low-dimensional
		functional of the component-specific parameters of the
		mixture. CliPS maps the component-specific MCMC draws to the point
		process representation and identifies clusters there, exploiting
		that, while data distributions usually overlap, the posterior of
		these functionals are more and more separated for increasing sample
		size. We outline the procedure and illustrate its use on several
		model-based clustering examples.\\[10pt]
		
		}

\maketitle	

	\section{Introduction}\label{sec:introduction}
	
	
	Cluster analysis is an unsupervised method to find groups in data and
	detect patterns as well as to reveal unobserved heterogeneity. Cluster
	analysis can be performed using algorithmic approaches, such as
	$k$-means \citep{Hartigan+Wong:1979}, or using a model-based approach
	\citep{Fraley+Raftery:2002}. In the model-based approach, mixture
	models are usually used with components assumed to correspond
	to clusters. A thorough and detailed introduction into finite
	mixture models including their use for clustering as well as
	semi-parametric density estimation is provided by
	\citet{McLachlan+Peel:2000} and \citet{Mix:Fruehwirth2006}.  A more
	recent overview and discussion of new developments in mixture analysis
	is provided in \citet{Fruehwirth-Schnatter:Celeux+Robert:2019}.
	
	\SFSblue{Model-based clustering applications arise in a range of
		different domains such as biology, business, economics, medicine,
		and psychology among others and involve different clustering
		kernels, i.e., the distributions or models used for the
		components. In this contribution, we present three different use
		cases from medicine, psychology and economics.  In the medical use
		case, three clinical metric measurements are available to infer
		different diabetes groups of patients and multivariate Gaussian
		distributions are used for the components as clustering kernels. In
		the psychological use case, multivariate categorical variables are
		available to characterize behavior of children with respect to their
		approach towards the unfamiliar and the clustering kernels are
		products of multinomial distributions assuming conditional
		independence of the categorical variables given cluster
		membership. The final economic use case aims at identifying groups
		of individuals with similar wage mobility behavior based on time
		series data about the income group an individual belongs to and uses
		first-order Markov chain models as clustering kernel.}
	
	The current contribution focuses on Bayesian model-based cluster
	analysis, see the pioneering papers of \cite{bin:bay},
	\cite{Fraley+Raftery:2002} and \citet{lau-gre:bay}.  Using the
	Bayesian framework to specify and estimate a mixture model for
	model-based clustering has several advantages: (1) Prior distributions
	allow to specify the prototypical shape of the clusters
	\citep{Hennig:2015}. This can be used to include domain knowledge in a
	principled way or to guide the estimation procedure to focus on
	mixture models with certain characteristics.  (2) Specifying priors
	enables the regularization and smoothing of the mixture likelihood.
	This is of particular importance, given the
	high irregularity of the mixture likelihood due to identifiability
	issues and multimodality as well as the presence of spurious modes.
	(3) Parameter uncertainty can be fully  assessed using the 
	corresponding
	posterior distribution.  No reliance on asymptotic normality is
	required, allowing for valid inference in cases where regularity
	conditions are violated, for small data sets and mixtures with
	small component weights.
	
	
	
	Our Bayesian model-based clustering approach is different from
	many Bayesian cluster analysis approaches where focus is on
	estimating the posterior of the partitions given a data set and 
	primary interest lies in determining a suitable partition of the
	observed data, see, e.g., \citet{wad-gha:bay}.  This partition then might be subsequently used to
	characterize the groups, see for example
	\citet{clips:Molitor+Papathomas+Jerrett:2010}. By contrast, we use the
	data to infer a statistical model which allows to determine the
	cluster distributions of the data generating process and thus
	characterizes the population of interest, beyond focusing only on the
	partition of the sample at hand. Our approach is more in line with
	usual statistical inference methods where the aim is to characterize
	the population and not only the data at hand. In many clustering
	applications, this perspective
	is also more natural, in particular, when a
	random or representative sample is analyzed to make inference about
	sub-populations present in the population. For instance, in marketing
	interest lies in identifying different consumer groups which may be
	characterized by their behavior rather than obtaining a partition of
	the consumers in the sample.
	
	Two aspects complicate the identification of the cluster
	distributions.  First, in many clustering applications the number of
	clusters is not known \textit{a priori}  and needs to be inferred from the data.
	Bayesian approaches either infer the number of  
	components in finite mixtures 
	via marginal likelihoods \citep{fru:est} or via  trans-dimensional posterior sampling \citep{Richardson+Green:1997}.
	In the present paper,
	we use an 
approach for fitting Bayesian mixture models
where a clear distinction is made between the number of components $K$ in the specified mixture \SFSblue{model} and the number of ``filled'' components $K_+$, which
eventually correspond to clusters.  We estimate the posterior
of filled components and thus infer a suitable number of clusters
based on this posterior, e.g., by using its mode as point estimate.

\SFSblue{Such a posterior estimate is obtained quite naturally using the framework of Bayesian mixture of
	finite mixture models where a prior 
	$p(K)$ is put on the unknown number of components $K$
	\citep{clips:Fruehwirth-Schnatter+Malsiner-Walli+Gruen:2021}. As discussed in \citet{gre-etal:spy}, 
	together with the prior on the \SFSblue{component} weights this implicitly induces a prior on the number of clusters  $K_+$ which differs from $p(K)$. 
	Alternatively, 
	a Bayesian non-parametric infinite mixture approach
	based on Dirichlet process
	mixture (DPM) models  \citep{clips:Kalli2011} 
	can be applied, provided that  a prior
	is placed
	on the concentration parameter.
	As recently 
	shown  in \citet{asc-etal:clu}, contrary to previous findings of inconsistency \citep{mil-har:sim}, 
	the number of clusters in a DPM
	model can be estimated consistently when such a prior is employed. A third strategy is to use a sparse finite mixture approach
	\citep{clips:Malsiner-Walli+Fruehwirth-Schnatter+Gruen:2016} where the number of clusters is inferred from an overfitting finite mixture model.    \citet{clips:Fruehwirth2019} indicate that this approach yields similar results as DPM, 
	once the prior on the concentration parameter is suitably matched to the prior on the component weights of the overfitting finite mixture.}

Pursuing either approach, Markov chain Monte Carlo (MCMC) sampling
with data augmentation for the missing cluster membership indicators
can be performed \SFSblue{to obtain draws from the posterior distribution}
\citep{Diebolt+Robert:1994}. These posterior draws are prone to label
switching and post-processing is necessary for cluster-specific
inference \citep{fru:mcm}. This represents the second issue when identifying the
cluster distributions, namely solving the label switching issue
\citep{Redner+Walker:1984}. Approaches to address the issue of label
switching and suitably post-process the posterior draws in Bayesian
cluster analysis differ by their focus on either solely obtaining a
consensus partition or, in contrast, relabeling the component-specific
parameters to obtain an identified mixture model. Methods focusing on
obtaining a final partition are based on the posterior of the
partitions which are label-invariant as they only represent if
observations are assigned to the same cluster or different
clusters. \citet{clips:Liverani+Hastie+Azizi:2015} use partitioning
around medoids with the posterior matrix of co-assignments transformed
from a similarity to a dissimilarity matrix as input to obtain a final
partition. Alternative approaches define a loss function between
partitions and obtain the partition which minimizes the expected posterior loss
\citep{clips:Dahl+Johnson+Mueller:2022}. Approaches using the
parameters of the mixture distribution were suggested by
\citet{clips:Stephens:2000, clips:Jasra+Holmes+Stephens:2005,
	Gruen+Leisch:2009, clips:Sperrin+Jaki+Wit:2010} among others and
several methods are implemented in the \textsf{R} \citep{clips:R:2025}
package \textbf{label.switching} \citep{clips:Papastamoulis:2016}. All
these approaches have in common that regardless of the suitability of
the fitted Bayesian mixture model for clustering, a final partition or
an identified model is obtained in any case, and no additional
information about the appropriateness of the inferred cluster
structure is provided.

Our approach by contrast does not only use a post-processing step to
solve the label switching problem in an automatic way but also
validates the obtained cluster structure and identifies the cluster
distributions. Conditional on a suitable number of clusters, the
proposed approach allows to assess how well the fitted mixture model
allows to infer the clusters and how the cluster-specific
distributions may be characterized.  This assessment relies on the
point process representation (\PPR) of a mixture model
\citep{clips:Stephens2000a}. The \PPR\ is employed to inspect the MCMC
draws of the sampled component-specific parameters or at least a
lower-dimensional functional thereof which is intended to characterize
the cluster distributions.  Our proposed approach is generically
applicable in a model-based clustering context to assess the cluster
structure and eventually obtain separated cluster distributions. The
approach may be used regardless of the component-specific
distributions or models used to characterize the clusters.
\SFSblue{In general, not all parameters of the component-specific
	distributions are needed to characterize the clusters and the user
	might even select a specific set or functional based on the
	clustering aims, i.e., depending on which characteristics are supposed to
	differentiate between clusters.}
Finally, the identified
mixture model allows to assign (new) observations to the estimated
clusters and thus to perform clustering not only on the available data,
but also for new observations.

The rest of the paper is organized as follows.
Section~\ref{sec:mixtures-known} introduces the Bayesian mixture
model, discusses the \PPR\ of a mixture model and outlines the
\emph{Clustering in the Parameter Space} (CliPS) procedure for
Bayesian model-based clustering in case the number of clusters is
known. Section~\ref{sec:mixtures-unknown} extends CliPS to the case
where the number of clusters is unknown and estimated through a
suitable specification of a Bayesian mixture model.  We demonstrate
the application of the CliPS procedure when the number of clusters is
unknown on empirical examples in Section~\ref{sec:case-studies}. Finally, we 
conclude in Section~\ref{sec:disc-concl} with a discussion of the applicability of the CliPS procedure beyond these specific frameworks.

\section{Mixtures 
	where
	the number of clusters is known}
\label{sec:mixtures-known}

\subsection{Model-based clustering via mixture models}

In the standard model-based clustering context, observations are
assumed to be drawn from a mixture distribution where each component
of the mixture corresponds to one data cluster and thus the number of
components and clusters coincide. This means that assuming $K$
clusters, 
the data $\ym=(\ym_1, \ldots, \ym_N)$ are a random sample, 
drawn from the following
distribution:
\begin{align}
	\ym_i &\sim \sum_{k=1}^K \eta_k f(\ym_i | \bm{\theta}_k),   \label{eq:mix}
\end{align}
where the component weights $\bm{\eta}_K = (\eta_1, \ldots, \eta_K)$
fulfill 
the conditions
$\eta_k > 0$, for $k=1,\ldots,K$, and $\sum_{k=1}^K \eta_k =1$,
and $f(\cdot | \bm{\theta}_k)$ represents the distribution of the
$k$-th component characterized by the $\betad$-dimensional
parameter vector
$\bm{\theta}_k=\trans{(\theta_{k1}, \ldots, \theta_{k
		\betad})}$. The component weights correspond to the cluster sizes
and the component distributions to the cluster distributions. 

A finite mixture distribution has a useful representation as a hierarchical generative model involving latent assignment variables
$(S_1,\ldots,S_N)$, where $S_i \in \{1, \ldots,K\}$ indicates the
component an observation $\by_i$ belongs to:
\begin{equation} \label{eq:mixhier}
	\begin{aligned}
		& S_i|K ,\bm{\eta}_K \sim \mathcal{M}_K(1, \bm{\eta}_K), \quad  \text{independently for } i=1, \ldots,N ,
		\\
		&  \ym_i | S_i = k, \Theta_K \sim  f(\ym_i | \bm{\theta}_k),   \end{aligned}
\end{equation}
where \SFSblue{$\mathcal{M}_K(1, \bm{\eta})$ is the $K$-dimensional
	multinomial distribution with one trial,
	$\bm{\eta}_K=(\eta_1, \ldots,\eta_K)$ is the vector of components
	weights and $\Theta_K = (\bm{\theta}_1, \ldots, \bm{\theta}_K)$ is
	the collection of component-specific parameters}.  This standard
finite mixture setting can be applied to model-based clustering of
very general data types such as \SFSblue{for example} uni- or multivariate data of
continuous and discrete variables.

\subsection{Bayesian finite mixture analysis}
\label{sec:BayMix}

In the Bayesian framework, all information contained in  
the data $\ym=(\ym_1, \ldots, \ym_N)$
about the parameters
$\bvartheta_K=(\boldeta_K,\btheta_1,\ldots,\btheta_K)$ of the mixture
model (\ref{eq:mix}) is summarized in terms of the posterior density
$p(\bvartheta_K|\by)$, which is derived by combining the mixture likelihood $p(\by|\bvartheta_K)$ with
a prior $p(\bvartheta_K)$ 
on the parameters 
using Bayes' theorem:
\begin{align*}
	p(\bvartheta_K|\by) &\propto  p(\by|\bvartheta_K)p(\bvartheta_K) = \SFSblue{\prod_{i=1}^N p(\by_i|\bvartheta_K)p(\bvartheta_K)}.
\end{align*}

\paragraph{Choosing the prior}

No conjugate prior exists for the mixture likelihood, thus the
following 
prior structure is usually assumed:
\begin{equation}
	p(\boldeta_K,\bTheta_K,\bzeta)   =p(\boldeta_K) p(\bTheta_K|\bzeta)p(\bzeta) = p(\boldeta_K) \prod_{k=1}^K p(\btheta_k|\bzeta)p(\bzeta), \label{eq:prior}
\end{equation}
where  $\bzeta$
are hyper-parameters which impact the component-specific prior 
distributions. This definition
assumes that the priors on the component weights
and on the component-specific parameters are independent and that the
component-specific parameters are conditionally
independent \textit{a priori}  given the \SFSblue{hyper-parameters $\bzeta$}. Usually, exchangeable
priors are employed 
which are invariant to permuting the component labels and
the same prior distribution results for each
component $k$, $k=1,\ldots,K$.
If, moreover, the priors on the \SFSblue{component} weights and component-specific
parameters are conditionally conjugate
in the hierarchical representation \eqref{eq:mixhier},
then the 
conditionally conjugate
posteriors of the
parameters are available in closed form. 

The conditionally conjugate prior for the components 
weights $\bm{\eta}_K=(\eta_1, \ldots,\eta_K)$ is the Dirichlet distribution. 
\SFSblue{To obtain a label-invariant
	prior, a symmetric $K$-dimensional Dirichlet distribution is specified,
	which depends only on a single scalar hyper-parameter $\gamma_K$:
	\begin{align} \label{priorDIR}
		\boldeta_K |K, \gamma_K
		\sim  \mathcal{D}
		(\gamma_K, \ldots, \gamma_K) = 
		\mathcal{D}_K(\gamma_K).
	\end{align}
	This prior is defined over the $K$-dimensional unit simplex and the choice of $\gamma_K$ determines prior concentration. The popular choice
	$\gamma_K=1$, e.g., implies a uniform distribution over all possible combinations of weights. 
	As $\gamma_K$ approaches 0, the prior more and more concentrates on the boundaries of the unit simplex, with
	some of the weights being very close to 0.}
The priors for the component-specific parameters
$\btheta_1,\ldots,\btheta_K$ depend on the component-specific
distribution of the mixture. They can be selected as the usual
conjugate priors of the component-specific distribution $f(\ym_i | \bm{\theta}_k)$, 
ignoring the
mixture structure. Regarding  the 
\SFSblue{hyper-parameter $\bzeta$}, 
\SFSblue{a  prior $p(\bzeta)$ is chosen which is 
	conjugate to  the 
	product  of all component-specific priors $\prod_{k=1}^K p(\btheta_k|\bzeta)$.} 

\paragraph{MCMC estimation}

For estimating the Bayesian finite mixture model and obtaining an
approximation of the posterior $p(\bvartheta_K|\by)$ different
computational solutions are available, see, e.g.,
\citet{Celeux+Kamaray+Malsiner-Walli:2019}. A generic approach builds
on data augmentation based on  the generative hierarchical model~\eqref{eq:mixhier} and includes the latent assignment variables
$\bS=(S_1,\ldots,S_N)$ 
in the sampling scheme
\citep{Diebolt+Robert:1994}.  
Using MCMC methods for simulation, posterior draws for both the unknown model parameters \SFSblue{$\bvartheta_K=(\boldeta_K,\bTheta_K,\bzeta)$, including all hyper-parameters $\bzeta$,} and the latent \SFSblue{assignment variables}
$\bS$ 
are sampled from the joint posterior $p(\bvartheta_K,\bS|\ym)$.

Under conditionally conjugate priors, the resulting MCMC
sampling scheme only requires Gibbs steps.
Given some initial parameter values $\boldeta_K^{(0)}$,  $\bTheta_K^{(0)}$ and \SFSblue{$\bzeta^{(0)}$}, the
following two steps are iterated for  $m = 1, \ldots, M_0 + M$:
\begin{enumerate}[1.]
	\item 
	Classification of each observation $\by_i$, $i = 1,\ldots,N$,  conditional on $\boldeta_K^{(m-1)}$ and $\bTheta_K^{(m-1)}$:
	\begin{enumerate}[(a)]
		\item 
		For $i = 1,\ldots,N$,  sample $S_i^{(m)}$ from the following discrete distribution:
		\begin{align*}
			\Prob{S_i = k|\by_i, \bvartheta_K^{(m-1)}} \propto \eta_k^{(m-1)} f(\by_i|\btheta_k^{(m-1)}), \quad 
			k=1, \ldots, K.
		\end{align*}
	\end{enumerate}
	\item Parameter simulation conditional on the classification $\bS^{(m)} = (S_1^{(m)},\ldots,S_N^{(m)})$:
	\begin{enumerate}[(a)]
		\item Sample \SFSblue{component} weights $\boldeta_K^{(m)}$ from
		\begin{align*}
			\boldeta_K|\bS^{(m)} &\sim   \mathcal{D} 
			(e_1,\ldots,e_K),
		\end{align*}
		where
		$e_k = \gamma_K + N_k$ with $N_k = \#\{i| S_i^{(m)} = k\}$ being    the number of observations assigned to component $k$.
		\item Sample the component-specific parameters $\btheta_k^{(m)}$, $k = 1,\ldots,K$,  conditional on $\by$, $\bS^{(m)}$ and $\bzeta^{(m-1)}$ from their conditional conjugate posteriors \SFSblue{$p(\btheta_k^{(m)}|\bS^{(m)},\bzeta^{(m-1)}, \by)$}.
		\item Sample the hyper-parameters $\bzeta^{(m)}$ conditional on $\bTheta_K^{(m)}$.
	\end{enumerate}
\end{enumerate}
\SFSblue{To remove the eventual bias resulting from the
	initialization}, the first
$M_0$ draws are discarded as burn-in and the remaining MCMC draws
$(\bvartheta_K^{(m)}, \bS^{(m)})$, $m=1,
\ldots,M$, are used for further inference.

The approximations of the posterior distributions of parameters and
latent variables should be identical across the $K$ components in case
the chain has converged and equally visited all modes induced by the
$K!$ permutations of the labels \citep{fru:mcm}. However, these approximations do not
help to characterize the component sizes and distributions or assign
observations to components. 

Hence, parameters which are
label-dependent need to be relabeled in order to focus on a single
mode of the posterior. Many different methods have been proposed to
obtain a unique labeling \citep[for an overview, see,
e.g.,][]{clips:Papastamoulis:2016}. For example, a unique labeling can
be achieved by imposing ordering constraints on the draws or by
solving an optimization problem to minimize the dissimilarity between
the classification matrices and some selected classification.
Alternatively, a straightforward approach is to exploit the  point process representation (\PPR) of
the MCMC draws to identify the mixture model
\citep{Mix:Fruehwirth2006,
	clips:Malsiner-Walli+Fruehwirth-Schnatter+Gruen:2016}, as will be
explained in detail in the next section.

\subsection{Point process representation of a mixture model}
\label{PPPRMix}

If  the component densities
$f(\cdot | \bm{\theta}_k)$ 
in the mixture model (\ref{ex:model}) arise from the same distribution family,
the clusters are characterized by the parameter vectors
$(\bm{\theta}_k)_{k=1,\ldots,K}$ of the underlying cluster
distribution. Clearly, these need to be pairwise different to ensure
identifiability. More specifically, any pair
$(\bm{\theta}_k,\bm{\theta}_{\knot})$ with $k \neq \knot$ needs to
differ in at least one component $\theta_{\cdot j}$, but not
necessarily in all of them, and these components need not be the same  for all pairs of clusters.  For a Gaussian mixture, for instance, 
$\bm{\theta}_k=\trans{(\mu_{k}, \sigma^2_{k})}$ and $f(\ym_i | \bm{\theta}_k)=f_N(\ym_i | \mu_k, \sigma^2_k)$ is equal to a  Gaussian density with mean
$\mu_k$ and variance $\sigma^2_k$. Some clusters could only differ in the mean $\mu_k$, some clusters only in the variance $ \sigma^2_k$, and some clusters in both elements. 

The full set of parameters which characterizes the mixture
distribution consists of
$\bvartheta_K = (\eta_k, \bm{\theta}_k)_{k=1,\ldots,K}$. The
parameterization, however, is not unique, because the labels can be
permuted.
A mathematical identification constraint characterizing the
differences between the cluster-specific parameters $\bm{\theta}_k$
could be used to obtain a unique labeling and thus an identified
parameterization of a mixture model.  \SFSblue{For a Gaussian mixture, for instance, where the clusters
	differ in the mean $\mu_k$, the ordering constraint  $\mu_1 < \ldots < \mu_K$ could be used.  
	However, if not all clusters  differ in the mean $\mu_k$ and  some clusters only differ in the variance $ \sigma^2_k$, such a simple ordering constraint  would not yield a unique labeling.  While mathematical identification can  be achieved also in this case, finding valid constraints is not straightforward, in  particular
	for  mixtures with higher dimensional parameters  $\{\bm{\theta}_1, \ldots, \bm{\theta}_K\}$, as will be illustrated below in Example~\ref{ex:model}.}


%

An alternative way to inspect a mixture model is to visualize its
parameters in the point process representation (\PPR),
a viewpoint introduced by \citet{clips:Stephens2000a}. The \SFSblue{use of the} \PPR\ was
further developed in \citet{Mix:Fruehwirth2006,
	clips:Fruehwirth-Schnatter:2011} with the aim of achieving
a unique labeling without the need to impose \textit{a priori}  any
mathematical identification constraints. 
The \PPR\ of a mixture model views the parameter vectors
$\bm{\theta}_k$ as points  and the component sizes
$\eta_k$ as corresponding marks for a marked 
point process. For fixed $K$ and a fixed
parametric family, any finite mixture distribution may be seen as a
distribution of the points $\{\btheta_1, \ldots , \btheta_K\}$ over
the parameter space, where each point $\btheta_k$ has an associated
positive mark $\eta_k$ and the marks are constrained to sum to one.

\paragraph{The \PPR\ for bivariate parameters}

The \PPR\ can be easily visualized for bivariate component-specific parameters $\btheta_k$. For illustration, 
the left-hand side of Figure~\ref{plot:PPR_mix_draws}
shows the \PPR\  of a mixture
consisting of three univariate Gaussian components. In particular, the
mixture model is given by
\begin{align*}
	y_i 
	\sim 0.3 f_N(
	y_i 
	| -3, 1) +
	0.5 f_N(
	y_i 
	| 0, 0.5) +
	0.2 f_N(y_i 
	| 2, 0.8).
\end{align*}
The parameter space thus is two-dimensional consisting of mean and
variance. The three bullets in Figure~\ref{plot:PPR_mix_draws} (left)
represent the \SFSblue{parameter vectors} of the three components. Evidently,
the \PPR\ is invariant to relabeling the components of the mixture
distribution and the mixture components differ with respect to mean as
well as variance. Also the different component sizes may be discerned
as indicated by the different sizes of the bullets.

\begin{figure}[t!]
	\centering
	\includegraphics[width=\textwidth]{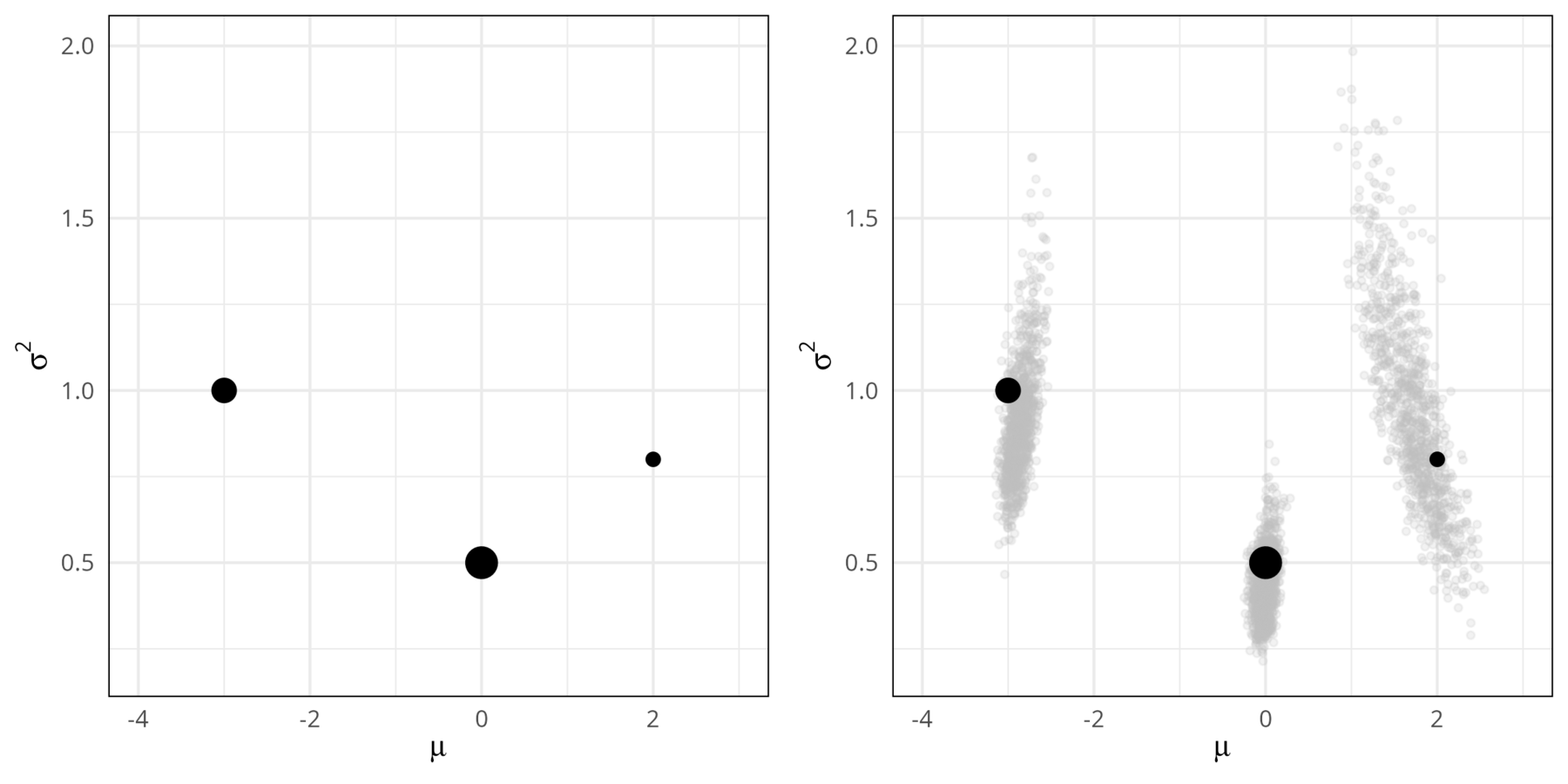}
	\caption{Left: \PPR\  of a univariate Gaussian mixture distribution
		with $K = 3$ components. Right: \PPR\  of the MCMC
		draws obtained when fitting a Gaussian mixture distribution with
		$K = 3$ components to 500 observations simulated from the mixture
		on the left.  } \label{plot:PPR_mix_draws}
\end{figure}


The \PPR\ has the advantage that it is a label-invariant
representation of the mixture, allowing us to highlight the
differences between the component distributions.
The \PPR\ is not only useful to visualize the parameterization of a given
mixture model, but it can also be used to visualize the posterior
distributions obtained through MCMC sampling by mapping each
single draw of the component-specific parameters to the \PPR\ and
plotting them together \citep{Mix:Fruehwirth2006,
	clips:Fruehwirth-Schnatter:2011}. In particular, if the fitted
mixture distribution matches the ``true'' mixture distribution, i.e.,
the component densities are correctly specified and the number of
components $K$ equals the number of generating components $K_+$
of
the sample, then the posterior draws of the component-specific
parameters are expected to cluster \SFSblue{for increasing sample size} around the ``true'' points
$\{\btheta_1^{\text{true}},\ldots,\btheta^{\text{true}}_{K}\}$ in
the \PPR , due to increasing posterior concentration.

The right-hand side of Figure~\ref{plot:PPR_mix_draws}
illustrates this using MCMC
draws from a mixture with three Gaussian components fitted to data
generated by the mixture model visualized on the left. In particular,
a sample of $N = 500$ observations
$\ym=(y_1, \ldots, y_N)$ is drawn from the specified mixture model. 
A Bayesian mixture model with $K = 3$ is fitted recording 1,000
MCMC iterations after omitting 1,000 burn-in iterations. For the prior
specification, see Example~\ref{ex:known} in Section~\ref{sec:CliPS-known-k_+}.

Rather than imposing \textit{a priori}  any artificial mathematical
identification constraints, the \PPR\ of the MCMC draws allows us
to study differences in the component-specific parameters
$\{\btheta_1, \ldots , \btheta_K\}$ that are supported by the
data \textit{a posteriori}.  This makes it very useful for identification of the cluster
distributions and for achieving a unique labeling.
The right-hand side of Figure~\ref{plot:PPR_mix_draws}, for
instance, clearly shows that the posterior draws of the parameters
concentrate on three well-separated clusters in the two-dimensional
parameter space and each draw of component-specific parameters can be
unambiguously assigned to one of the three groups. This assignment can
be used to label the components and identify the cluster
distributions. It is also obvious that the posterior distributions are
more diffuse for components with a small cluster weight
$\eta_k$. As fewer observations are generated by such a cluster,
less information in the data is available to learn the
corresponding component-specific parameter $\btheta_k$.

While the MCMC draws projected to the two-dimensional
parameter space clearly indicate that the three component-specific
parameters are different, the differences in any single dimension
are less pronounced for identification. In particular, achieving a
unique labeling based solely on the ordering constraint
$\sigma_1^2 < \ldots < \sigma_K^2$ on the variances would be
impossible for the given data set  
(although, of course, since the true variances are different, for a much larger
sample size $N$, posterior concentration would increase and it
might be possible to use this constraint for identification).


\paragraph{Exploring the  \PPR\ for multivariate parameters}

The indicated identification of the cluster distributions
from a visualization of the \PPR\ is straightforward only in a
two-dimensional parameter space. While a \PPR\ clearly also exists for
multivariate parameters of dimension $\betad >2$, exploring the
\PPR\ in the $\betad$-dimensional space is as challenging as
exploring multivariate data.  For visualization, scatter
plots of any of two components $\theta_{\cdot j}$ and
$\theta_{\cdot \ell}$ of $\bm{\theta}_{.}$ with $j \neq \ell$, similar to
Figure~\ref{plot:PPR_mix_draws}, can be used to produce pairwise
\PPR s.  Obviously there are $\betad (\betad-1)/2$ such pairwise
\PPR s, but we need not consider all of them for visualization, in
particular for large $\betad$. The goal is to explore in which sense
the parameters differ across the clusters and we may focus on those
components which we expect to be different based on
\SFSblue{domain-specific} 
knowledge
about the observed data.


Note that also a lower-dimensional functional
$\funcTh{\bm{\theta}_k}$ of the component-specific parameter
vectors $\bm{\theta}_k$ can be used for visualization. This functional
should reflect the aspect in which clusters are assumed to
differ. E.g., in many model-based clustering applications where a
mixture of multivariate Gaussians 
with component-specific distributions $\ym_i|S_i=k \sim  \mathcal{N}_r(\bm{\mu}_k,\bm{\Sigma}_k)$
is fitted to
$r$-dimensional data, the component-specific mean vectors $\bmu_k$
characterize the clusters, whereas the 
component-specific covariance
matrices $\bm{\Sigma}_k$
represent nuisance parameters.
Hence, the \PPR\  could focus on 
exploration 
of the $r$-dimensional functional  $\funcTh{\bm{\theta}_k}=\bmu_k$ instead of exploring  all $r+r(r+1)/2$ distinct values in $\btheta_k=(\bmu_k, \bSigma_k)$.

For illustration, we consider an artificial example where the data
generating process corresponds to a  multivariate Gaussian
mixture with four components.  The four components correspond to
four clusters to be inferred. The cluster distributions differ in
the mean values of at least one 
dimension, however,  
the component-specific covariance
matrices $\bm{\Sigma}_k$ are identical across all clusters.
\SFSblue{The mean parameters thus characterize the cluster
distributions.} We will use this example in this section to
illustrate the \PPR\  of the draws of a mixture model when the number of
clusters is known 
as well as in Section~\ref{sec:mixtures-unknown} when the number of
clusters has to be estimated.

\begin{example}[{\bf Illustrative example}]\label{ex:model}
We consider a mixture of $K = 4$ multivariate Gaussian distributions
of dimension $r = 6$ with 
$\eta_1=\ldots=\eta_4=1/4$,   
\begin{align} \label{defineGM}
	(\bm{\mu})_{k=1,\ldots, 4 
	} &= \left(
	\begin{array}{rrrr}
		-2&-2&-2&2\\
		-3&3&-3&3\\
		4&4&4&4\\
		0&0&0&0\\
		2&2&0&2\\
		2&0&0&2
	\end{array}\right), \qquad  \qquad 
\end{align}
and $ \bm{\Sigma}_k = 0.6 \bm{I}_6$, for all $k=1,\ldots, 4$,
where $\bI_r$ is the $(r \times r)$ identity matrix.  The
component-specific parameter vector $\btheta_k=(\bmu_k, \bSigma_k)$
is of dimension $r + r(r + 1)/2 = 27$. Clearly in this mixture model
the components only differ in their mean parameters, but not their covariance matrices.

Figure~\ref{plot:example-fig1} (left) shows the pairwise \PPR\  of this mixture,
using only 
the mean parameters  in the various dimensions for visualization.
The variance matrices are
the same across components and do not provide any information to
differentiate between the components. 
\SFSblue{While the mean parameter vectors 
	$\bm{\mu}_1, \ldots, \bm{\mu}_K$ defined
	in \eqref{defineGM}
	are distinct, not a single dimension of $\bm{\mu}_k$ is different across all four components.
	Moreover,  in two dimensions  the $\bm{\mu}_k$s do not differ at all. }  
Consequently, only a single plot out of the 15  pairwise \PPR s 
(i.e., dimensions two and six)
shows four bullets. The less heterogeneous the means are, the fewer bullets are present. In one specific pairwise \PPR\ (i.e., dimensions three and four)
only a single bullet is visible due to  homogeneity of the means along these two dimensions.

Figure~\ref{plot:example-fig1} (right) shows pairwise scatter plots of 1,000 observations drawn from the specified mixture model.
\SFSblue{This data set 
	will be used later in Example~\ref{ex:known} to illustrate the CliPS procedure.} 
If a Bayesian mixture model with $K=4$ components is fitted to these data,   MCMC sampling will provide us with $M$ posterior draws $\bmu ^{(m)} = (\bmu_{1} ^{(m)}, \ldots, \bmu_{4} ^{(m)} )$, $m=1, \ldots, M$.    
Similarly as the MCMC draws of a bivariate parameter shown in
Figure~\ref{plot:PPR_mix_draws}, a pairwise \PPR\ of these
posterior draws -- shown in Figure~\ref{plot:example-fig2} --
mirrors the structure of the true \PPR\ provided on the left hand
of Figure~\ref{plot:example-fig1}. More generally, when we
consider the MCMC draw
$\bmu_{k} ^{(m)}, k=1,\ldots,4, m=1, \ldots,M$, as points in the
$r$-dimensional parameter space, they will cluster around the
four points defined by the true mean vectors in
\eqref{defineGM}. Hence, we can apply any unsupervised clustering
method suitable for multivariate data to the $4M$ posterior draws
$\{\bmu_{k} ^{(m)}\}$ to implicitly identify which component-specific MCMC draws
belong to the same mixture components. This idea of \lq\lq
clustering in the parameter space\rq\rq\ (CliPS) to identify the
components is described in a generic manner in
Section~\ref{sec:CliPS-known-k_+}.
\end{example}

\begin{figure}[t!]
\centering
\includegraphics[width=0.49\textwidth]{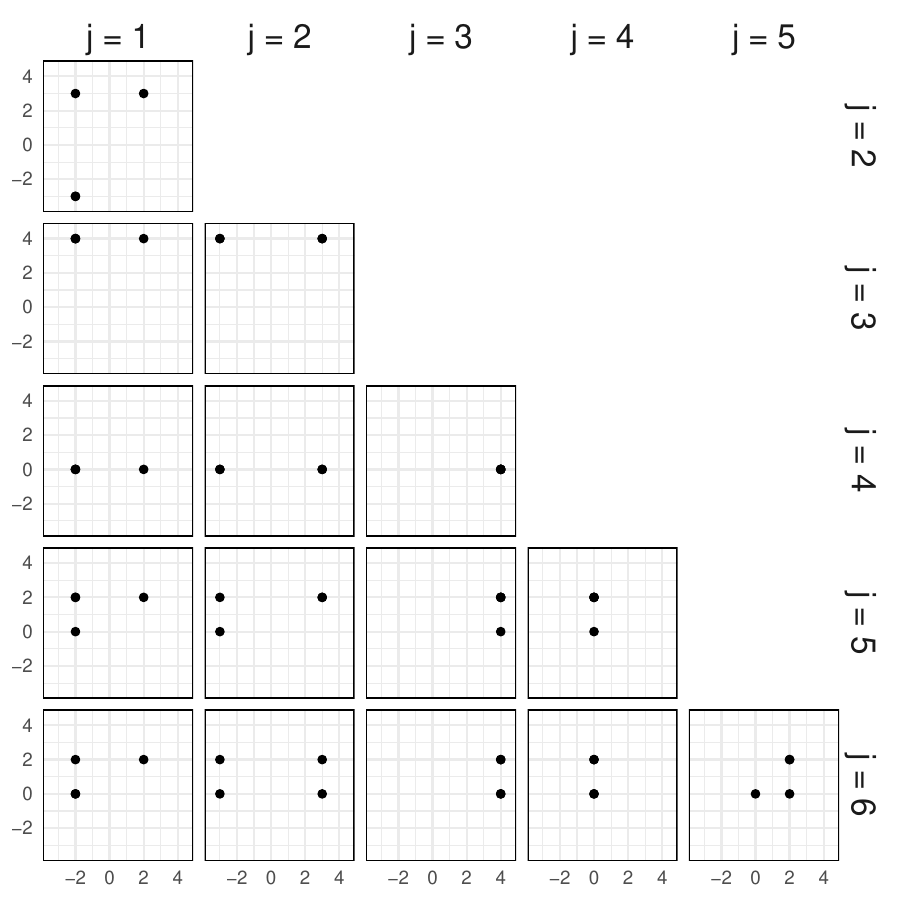}
\includegraphics[width=0.49\textwidth]{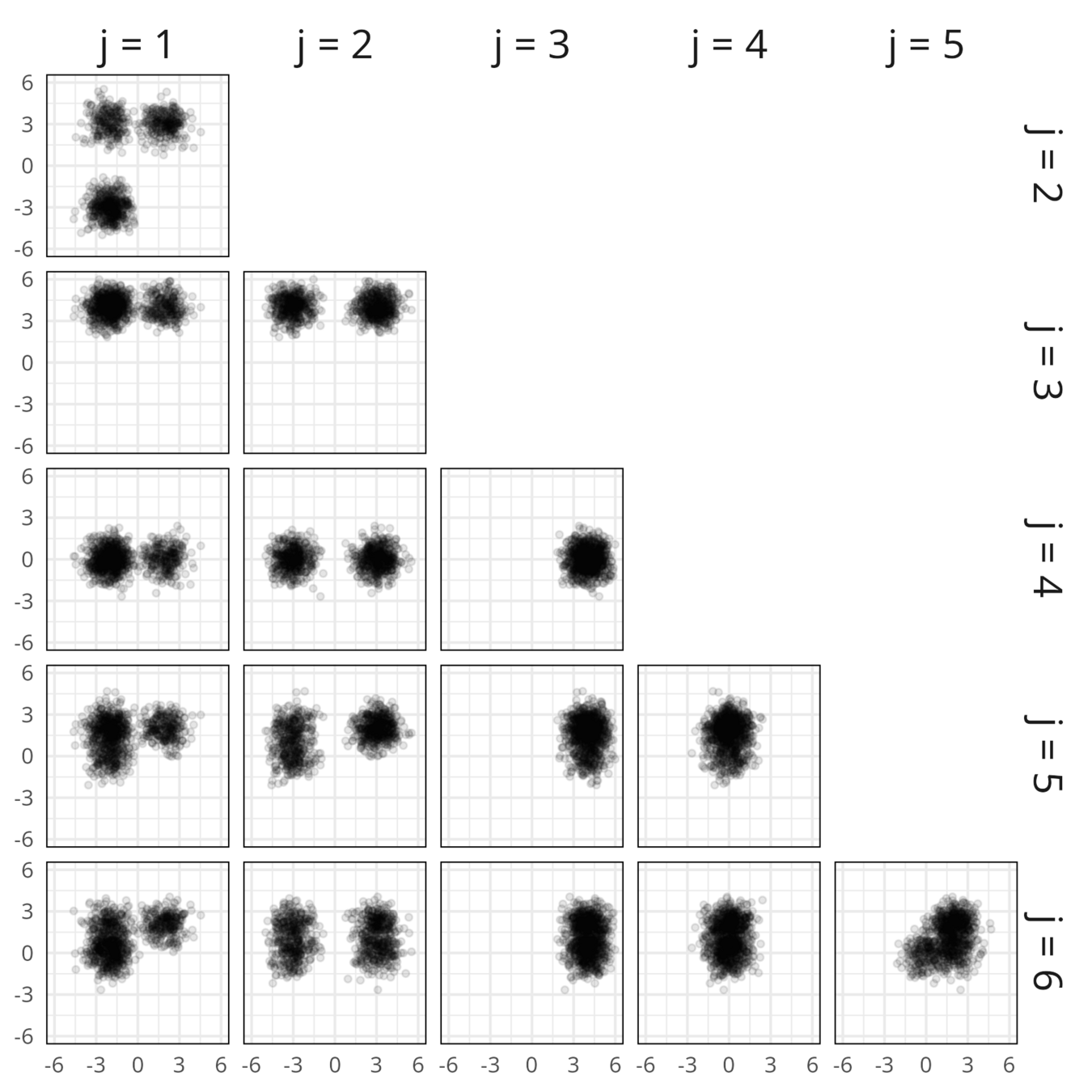}
\caption{{\bf Example 1}. Left: pairwise \PPR\ of 
	a six-dimensional Gaussian mixture model with $K = 4$
	clusters using 
	the mean parameters in the various dimensions for
	visualization. Right: corresponding pairwise scatter
	plots of 1,000 observations drawn from the mixture model on the
	left.}
\label{plot:example-fig1}
\end{figure}


\subsection{The CliPS procedure when the number of clusters is known}\label{sec:CliPS-known-k_+}

In the following we assume that the number of clusters $K_+$ in a given data set $\ym=(\ym_1, \ldots, \ym_N)$ is
known. Then the workflow of the  CliPS procedure basically consists of 
the following steps: (i) specifying 
a Bayesian mixture
model with $K=K_+$  components   
as in Section~\ref{sec:BayMix}, consisting of the data model as well as the prior
specifications which reflect the prior knowledge about the cluster
shapes, (ii) fitting the model to the data
using MCMC sampling, (iii) clustering 
the sampled
parameters in the \PPR\  and assessing the suitability of the cluster
solution and (iv) obtaining an identified model by suitable relabeling. 
More specifically, 
the procedure involves the following steps:
\begin{enumerate}[Step 1.]
\item 
Define a Bayesian mixture model with $K=K_+$ components, 
by combining the generative hierarchical model \eqref{eq:mixhier}
with the prior distribution \eqref{eq:prior}.
\item Use MCMC sampling with data augmentation as described in Section~\ref{sec:BayMix} to simulate 
all unknown model
parameters 
\SFSblue{$\bvartheta_K = (\boldeta_K,\btheta_1, \ldots, \btheta_K, \bzeta)$}
and latent \SFSblue{assignment variables} 
$\bS=(S_1, \ldots, S_N)$ from the posterior
distribution \SFSblue{$p(\bvartheta_K|\ym)$}. Record $M$ iterations after omitting 
$M_0$ burn-in iterations and potentially applying thinning to
an MCMC chain of size $M_0 + C M$ with $C\geq 1$.
\item Arrange the component-specific parameter draws $\btheta_k^{(m)}$
(or a lower-dimensional functional $\funcTh{\btheta_k^{(m)}}$),
$m=1,\ldots,M$, for each $k$ on top of each other, resulting in a
matrix with $KM$ rows and $d$ columns, where
$d=\dim(\btheta_k^{(m)})$ (or $d = \dim(\funcTh{\btheta_k^{(m)}})$).
Cluster the $K M$ draws into $K$ groups using a clustering
algorithm, e.g., $k$-means clustering, to obtain a group assignment
$I_k^{(m)}$ of the $k$-th component of the $m$-th draw.
\item For each iteration $m$, $m = 1,\ldots,M$, construct the
classification sequence $\rho_m = (I_1^{(m)},\ldots,I_K^{(m)})$, and 
check whether $\rho_m$ defines 
a permutation of $\{1,\ldots,K\}$, e.g.,
by simply sorting the elements of $\rho_m$ and checking whether the
resulting sequence is equal to $\{1,\ldots,K\}$.
\item Determine the number $M_\nu$ of classification sequences
$(\rho_1, \ldots,  \rho_M)$  not being a permutation and the corresponding 
``non-permutation rate''
$\nu = M_\nu/M$.  Remove the $M_\nu$ iterations, where the
classification sequences $\rho_m$ are not permutations
of $\{1,\ldots,K\}$. For the remaining $M-M_\nu$ MCMC iterations, \SFSblue{the sequence
	$\rho_m$ is a permutation of $\{1,\ldots,K\}$}, and a unique labeling is achieved by
reordering the draws through the inverse $\rho_m^{-1}$ of $\rho_m$:
\begin{enumerate}[(1)]
	\item 
	$\eta_1^{(m)},\ldots,\eta_K^{(m)}$
	is substituted by
	$\eta^{(m)} _{\rho_m^{-1}(1)},\ldots,\eta^{(m)}_{\rho_m^{-1}(K)}$;
	\item $\btheta^{(m)}_1,\ldots,\btheta^{(m)}_K$ is substituted by
	$\btheta^{(m)}_{\rho_m^{-1}(1)},\ldots,\btheta^{(m)}_{\rho_m^{-1}(K)}$;
	\item $S_1^{(m)}, \ldots,S_N^{(m)}$ is substituted by
	$\rho_m^{-1}(S_1^{(m)})$, \ldots,
	$\rho_m^{-1}(S_N^{(m)})$.
\end{enumerate}
\item Use the relabeled draws to identify and characterize the cluster
distributions.
\end{enumerate}
Note that in Step~3, the information that the draws $\btheta_k^{(m)}$
and $\btheta_{k'}^{(m)}$ are from the same MCMC iteration is not used in
the clustering procedure and draws from different components $k$ and
$k'$ may be allocated to the same group. This is different from other
relabeling methods which force $\btheta_k^{(m)}$ and
$\btheta_{k'}^{(m)}$ to belong to different groups, in order to
guarantee that $\rho_m$ is a permutation for all draws.  This
``non-enforcing'' of permutations is a strength of the CliPS
approach, as this allows us to assess, whether the posterior draws,
clustering around the underlying parameter values, are separated
well enough to unambiguously be allocated to different groups. Only in this
case, the corresponding cluster distributions are well-separated and
can be uniquely identified.

The non-permutation rate $\nu$ obtained in Step~5 hence is a
crucial performance criterion indicating the suitability of the fitted
mixture model with the chosen number $K$ of components
as a cluster solution.  A low non-permutation rate
indicates that a unique characterization of the cluster distributions
is possible.

As motivated for a multivariate Gaussian mixture in Example~1
in Section~\ref{PPPRMix}, a lower-dimensional functional
$\funcTh{\btheta_k}$ of the component-specific parameter $\btheta_k$
can be used in Step~3.  Regardless of the chosen functional, the
classification sequences $\rho_m$, $m=1, \ldots,M$, resulting from the
clustering procedure carry all information we need to relabel the
MCMC draws of the entire component-specific parameters
$\{\btheta_1, \ldots, \btheta_K\}$, the weights   $\{\eta_1, \ldots, \eta_K\}$ and the class indicators
$(S_1, \ldots, S_N)$ in Step~5.

To illustrate Step~4 in more detail,   
consider a mixture with $K=4$, \SFSblue{where 
Step~3 is based on the entire parameter $\btheta_k$, i.e.,  $\funcTh{\btheta_k}=\btheta_k$.
Assume that the clustering procedure yields  the  classification sequence $\rho_m=(1,3,4,2)$ for the $m$-th posterior draw.} That
means that the draw of the first component was assigned to group one,
the draw of the second component to group three and the draw of the
third and fourth component to group four and two, respectively.  Thus,
the draws of this iteration are assigned to different groups in the
\PPR , which allows to relabel 
them. \SFSblue{Hence, $\rho_m=(1,3,4,2)$
indicates that
$\{\btheta_1 ^{(m)},\btheta_2 ^{(m)},\btheta_3 ^{(m)},\btheta_4^{(m)}\}$ 
are draws of the component-specific parameters 
$\{\btheta _1,\btheta_3,\btheta_4,\btheta_2\}$. 
To obtain a draw of $\{\btheta _1,\btheta_2,\btheta_3,\btheta_4\}$ under a  unique labeling,  
we need to relabel (rearrange) the original draws as $\{\btheta_1 ^{(m)},\btheta_4 ^{(m)},\btheta_2 ^{(m)},\btheta_3^{(m)}\}$.} 
Formally, the indices under a unique labeling are given by $(1,4,2,3)=(\rho_m^{-1}(1),\rho_m^{-1}(2),\rho_m^{-1}(3),\rho_m^{-1}(4))$.
However, if the classification
sequence $\rho_m=(2,1,2,3)$ is obtained for iteration $m$,
which is not a permutation of $\{1,\ldots,4\}$, 
then draws
sampled from two different components (component 1 and 3) are assigned
to the same group (group 2) and no relabeling can be performed for
this iteration. Thus the draws of this iteration have to be removed
from further inference.

All classification sequences $\rho_m$, $m=1,\dots,M$, are expected to
be permutations, iff the \PPR\  of the MCMC draws contains $K$
well-separated simulation clusters \SFSblue{in the parameter space}. If a small fraction of
non-permutations is present, then the posterior draws may be stratified
according to whether they can be identified or not. Only the
sub-sequence of identified draws obtained in Step~4 is used to relabel
the draws in Step~5 and in Step~6 for cluster-specific inference.
However, if the fraction of non-permutations is high for the chosen
number of components $K$, this indicates that posterior draws
generated from different components are strongly overlapping. The
resulting cluster distributions are thus not clearly separable.  A
high fraction of non-permutations typically happens in cases, where
the mixture is overfitting the number of cluster distributions. This
suggests that the fitted mixture model does not represent a suitable
cluster solution and reformulating the data model together with the
prior specification should be considered to assess if a better suited
cluster solution might be obtained.

The CliPS procedure requires clustering of the MCMC draws of the
component-specific parameters in the parameter space. 
If a 
suitable cluster solution is obtained, the cluster distributions should differ at
least with respect to some parameters in a sufficiently pronounced way
in order to allow their unique characterization through these
parameters. 

\paragraph{An illustrative example}

In the following we apply CliPS to the artificial data simulated in Example~1
from a mixture
of four multivariate Gaussian component distributions, assuming that
the number of clusters is known. We start by specifying a suitable
mixture model and obtain posterior draws using MCMC sampling with data
augmentation. We then use the \PPR\  of the mean parameters to assess the
suitability of the cluster solution, 
identify the model and
characterize the cluster distributions.

\begin{example}[\bf Known number of clusters]\label{ex:known}
We fit a Bayesian finite mixture model to a data set with
$N = 1,000$ observations from the data generating process given in
Example~\ref{ex:model}, which is visualized in
Figure~\ref{plot:example-fig1}. The following generative model is
assumed when fitting the model:
\begin{equation} \label{hierKknown}
	\begin{aligned}
		\bm{\eta}_K |K, \gamma_K &\sim \mathcal{D}_K(\gamma_K),\\
		\bm{C}_0 &\sim \mathcal{W}_r(g_0, \bm{G}_0),\\
		\bm{\mu}_k &\sim \mathcal{N}_r(\bm{b}_0, \bm{B}_0), \quad k=1, \ldots, K,\\
		\bm{\Sigma}_k &\sim \mathcal{W}_r^{-1}(c_0, \bm{C}_0), \quad  k=1, \ldots, K,\\
		S_i|K ,\bm{\eta}_K &\sim \mathcal{M}_K(1, \bm{\eta}_K), \quad  \text{independently for } i=1, \ldots,N ,\\
		\ym_i|S_i=k,\bm{\Theta}_K&\sim \mathcal{N}_r(\bm{\mu}_k,\bm{\Sigma}_k), \quad \text{independently for } i=1, \ldots,N,
	\end{aligned}
\end{equation}
where 
$\mathcal{W}_r(\alpha, \bmWi)$ is 
the
$r$-dimensional Wishart distribution with parameters $\alpha$ and
$\bmWi$ such that the mean is given by $\alpha \bmWi^{-1}$,
$\mathcal{N}_r(\bm{\mu}, \bm{\Sigma})$ the $r$-dimensional Gaussian
distribution with mean $\bm{\mu}$ and variance $\bm{\Sigma}$, and
$\mathcal{W}_r^{-1}(\alpha, \bmWi)$ the $r$-dimensional inverse
Wishart distribution with parameters $\alpha$ and $\bmWi$ such that
the mean is given by $\bmWi / (\alpha - (r+1)/2)$.

We draw MCMC samples with data augmentation using 
\SFSblue{the \textsf{R} package \textbf{telescope}
	\citep{clips:Malsiner-Walli+Gruen+Fruehwirth-Schnatter:2025},
	employing the following data-based
	hyper-parameters for the component-specific priors 
	\citep{clips:Fruehwirth-Schnatter+Malsiner-Walli+Gruen:2021}}:   $\bb_0=\text{median}(\ym)$, $\bB_0=\text{diag}(\text{range}(\ym))$,
$c_0 = 2.5 + (r-1)/2$,
$g_0 = 0.5 + (r-1)/2$,
$G_0 = 100 \cdot g_0/c_0 \cdot \bR^{-1}$, where $\bR=\text{diag}(R_1^2,\ldots,R_r^2)$ and $R_j$ is the range of the data in dimension $j$.
\SFSblue{Regarding the Dirichlet prior 
	on the weights, we choose $\gamma_K = 4$,
	as recommended by \citet{Mix:Fruehwirth2006} for mixtures where $K$ is known.} 
We
initialize the MCMC sampler by first determining a clustering of the
data points into four 
groups based on $k$-means clustering and
determining initial component-specific mean vectors and 
covariance matrices
based on the group-specific empirical means and covariances.   
The
component weights 
are initialized assuming equal sizes, i.e., $1/4$.

The pairwise \PPR\  of the various components $\mu_{\cdot j}$ of the mean parameters of 1,000 MCMC draws after omitting
the first 1,000 draws as burn-in is shown in
Figure~\ref{plot:example-fig2}. The gray points represent the MCMC
draws while the black bullets indicate the true parameter values. A
clear clustering of the MCMC draws around the true values is
visible. Compared to the data points as shown in the scatter plot in
Figure~\ref{plot:example-fig1}, identifying the clusters is much
easier \SFSblue{in this space than in the sample space}.  Clustering the mean parameters using $k$-means results in
group labels $\rho_m$ \SFSblue{which correspond  to permutations of $\{1,\ldots,4\}$} for all MCMC
draws, i.e., the non-permutation rate $\nu$ is equal to zero, indicating that a
suitable cluster solution was inferred with the cluster-specific
posterior distributions of the mean parameters being clearly
separable. All draws thus can be retained, relabeled and used for
posterior inference on the cluster distributions.
\end{example}


\begin{figure}[t!]
\centering
\includegraphics[width=0.49\textwidth]
{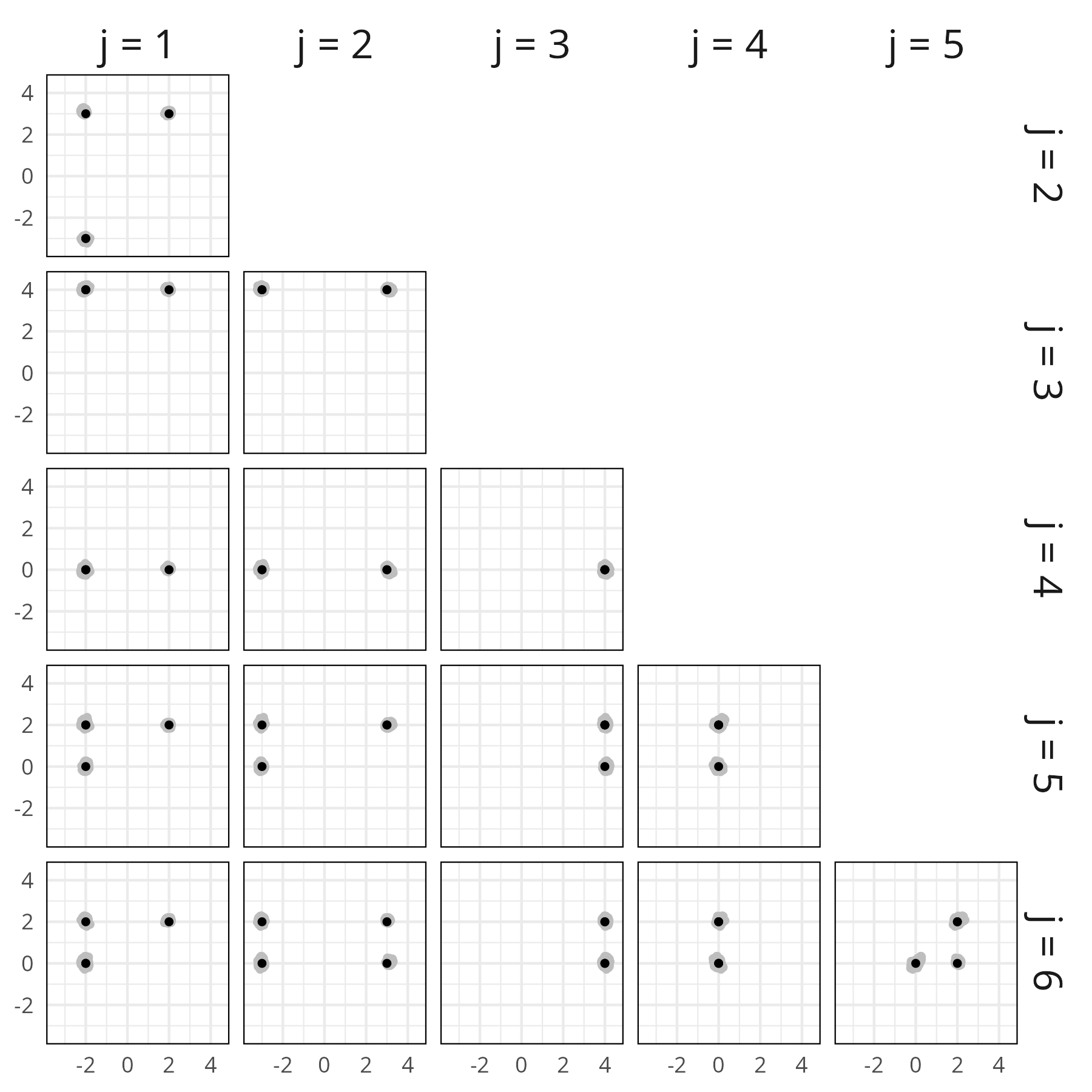}
\caption{{\bf Example 2} ($K = 4$ known).
	Pairwise \PPR\ of the MCMC draws using only the mean
	parameters in the various dimensions in gray together
	with the component means of the true mixture model
	indicated by black bullets.}\label{plot:example-fig2}
	\end{figure}
	

	\section{Mixtures 
where
the number of clusters is unknown}
\label{sec:mixtures-unknown}

\subsection{Bayesian mixture analysis with a prior on $K$}
\label{sec:BayKunknon}

In many applications, the number of clusters in the data is unknown
and has to be estimated. To this aim, a broad range of methods have been suggested, see
\cite{cel-etal:mod} for a recent overview of both classical and Bayesian techniques.

In Bayesian analysis, a natural approach to
account for the unknown number of components $K$ is to introduce a
prior on $K$ and 
to perform inference both on the
posterior of $K$ as
well as the posterior of $K_+$, the number of filled components which generated the clusters in the sample. 
The
hierarchical generative model given in \eqref{eq:mixhier} and
\eqref{eq:prior} 
then has  an 
additional layer at the
top:
\begin{align*}
K \sim p(K),  
\end{align*}
where $p(K)$ is the prior on the number of components. Following
\citet{Miller+Harrison:2018},
we call the resulting model a mixture of
finite mixtures (MFM) model.

An important distinction to make when using a prior on $K$ for estimating the number of
clusters $K_+$ in the data, is the conceptual difference between the number of
components $K$ in the finite mixture model and the number of clusters
$K_+$
in the data set.  \citet{McCullagh+Yang:2008}
discuss this distinction in the context of species sampling where $K$
represents the number of species in the population whereas $K_+$
corresponds to the number of species observed in the available data
set. These numbers need not to be the same as not all species of the
population might also be observed in the sample. Correspondingly, not
all available mixture components might be used to generate the data at
hand. Thus, it might happen that $K_+ < K$. In this case, redundant components
appear in the mixture distribution, indicating an ``overfitting''
mixture.

Specific identifiability issues arise in the MFM model as the handling of an overfitting mixture model where the
number of mixture components $K$ exceeds the number of clusters $K_+$
in the data set is not straightforward.  Additional components can be added to any mixture model without
changing the induced mixture distribution: either a component weight
of zero is assigned to the new components or the new components are
obtained by duplicating an already existing component and splitting
the component weight between them.

For Bayesian mixtures, \citet{Mix:RousseauMengerson2011} provide theoretic insights into how
the value $\gamma_K$ of the Dirichlet prior on the \SFSblue{component} weights
defined in \eqref{priorDIR}
impacts on the asymptotic behavior of the posterior distribution of an overfitting
finite mixture model. \SFSblue{They show 
that empty rather than redundant components are  encouraged asymptotically in
overfitting mixtures by choosing a sufficiently  small hyper-parameter $\gamma_K$ 
for the
Dirichlet prior on the weights. According to
\citet{clips:Malsiner-Walli+Fruehwirth-Schnatter+Gruen:2016},  in applied mixture analysis, 
$\gamma_K$ has to be 
much smaller than these asymptotic bounds to achieve the desired emptying of redundant components. Consequently, they call a finite mixture model where   $\gamma_K\ll 1$ a sparse finite mixture model.}

These results suggest the following strategy to infer the number of
clusters in Bayesian estimation in case the number of clusters is not
known: Select a prior on the number of components $K$ in the mixture
distribution to ensure that this number is larger than the number of
clusters $K_+$ in the data. In addition, choose parameter values of
the Dirichlet prior on the component weights to ensure that redundant
components are  left empty rather than being duplicated in case of an
overfitting mixture model. This then allows to decide on a suitable
number of clusters based on the posterior of the number of filled
components, see
\citet{clips:Malsiner-Walli+Fruehwirth-Schnatter+Gruen:2016} and 
\citet{clips:Fruehwirth-Schnatter+Malsiner-Walli+Gruen:2021}.

To this aim,
\citet{clips:Fruehwirth-Schnatter+Malsiner-Walli+Gruen:2021} propose a
suitable prior structure for a MFM model in the
context of model-based clustering. They propose to use
\begin{align}
K - 1 &\sim \mathcal{BNB}(1, 4, 3),\label{eq:BNB}
\end{align}
where $\mathcal{BNB}(a, b, c)$ denotes the beta-negative-binomial
distribution with mean 
\SFSblue{$a {c}/{(b-1)}$}.  
In addition, a so-called
dynamic specification of a MFM model is used in \citet{clips:Fruehwirth-Schnatter+Malsiner-Walli+Gruen:2021} 
for the prior of the weights:
\begin{align} \label{eq:weights}
\bm{\eta}_K |K, \gamma &\sim \mathcal{D}_K(\gamma / K), 
\end{align}  
\SFSblue{where the hyper-parameter $\gamma_K=\gamma / K$ is the smaller 
the
larger $K$, inducing in this way an increasingly sparse solution as $K$ increases and implying
a larger gap between $K$ and $K_+$ for larger values of $K$.
It should be emphasized that the  term \lq\lq dynamic\rq\rq\ in the context of  a MFM model refers to this specific variation
of the hyper-parameter $\gamma_K$ with respect to $K$ and does not indicate any kind of time-varying mixture model. 
We will also use
the priors in
\eqref{eq:BNB} and \eqref{eq:weights} in the subsequent examples where we fit a MFM 
model.}


\SFSblue{Various algorithms have been proposed to fit a MFM model, including reversible jump MCMC (RJMCMC; \cite{Richardson+Green:1997}),
birth-and-death sampling \citep{clips:Stephens2000a},
a finite version of the Chinese restaurant process (CRP) sampler  \citep{jai-nea:spl_2004, jai-nea:spl_2007} suggested
by \citet{Miller+Harrison:2018}  and the  telescoping sampler  
\citep{clips:Fruehwirth-Schnatter+Malsiner-Walli+Gruen:2021}.
The telescoping sampler   extends the 
MCMC scheme with data augmentation 
outlined in Section~\ref{sec:BayMix} to a setting where $K$ is unknown}
by including two further steps: 
a third step
where $K$ is explicitly sampled conditional on the current partition
of the data containing $K_+$ non-empty components and 
a fourth step
to 
sample component-specific parameters
for the empty components:
\begin{enumerate}
\item[3.] Sample $K$ conditional on current cluster sizes $(N_1,\ldots,N_{K_+})$ and $\gamma_K$ from
\begin{eqnarray*}
	p(K|N_1,\ldots,N_{K_+},\gamma_K) &\propto
	\frac{K!}{(K-K_+)!}
	\frac{\Gamma(K\gamma_K)}{\Gamma(K \gamma_K +N)}	 
	\prod_{k=1}^{K_+} \frac{\Gamma(N_k+ \gamma_K)}{\Gamma(1+ \gamma_K)} p(K).
\end{eqnarray*}
\item[4.] Add $K-K_+$ empty components with component-specific parameters
sampled from the priors.
\end{enumerate}
Note that in Step~2 of the 
MCMC scheme outlined in Section~\ref{sec:BayMix}, 
the \SFSblue{component} weights $\bm{\eta}_K$
are drawn in Step~(a) including the empty
components,
while the hyper-parameters are drawn in Step~(c)
conditional on the
component-specific parameters of the filled components, disregarding
the empty components.  
In contrast to
the RJMCMC sampler of
\citet{Richardson+Green:1997},
this sampling scheme is very generic, allowing for arbitrary component
distributions by just re-using sampling schemes already available for
finite mixtures with a fixed number of components $K$. In contrast to
the CRP sampler of
\citet{Miller+Harrison:2018}, which requires a constant hyper-parameter $\gamma_K=\gamma$, the
telescoping sampler is also applicable with arbitrary values selected
for $\gamma_K$ \SFSblue{and works particularly well for the dynamic MFM model, where  $\gamma_K=\gamma/K$.}

\SFSblue{Formally,} the MFM model encompasses the
sparse finite mixture model
\citep{clips:Malsiner-Walli+Fruehwirth-Schnatter+Gruen:2016} and the
Dirichlet process prior \citep{clips:Kalli2011} as special cases,
where the prior puts all mass on a fixed value being, respectively, 
equal to a large, but finite value ($K< \infty$) or infinite ($K= \infty$).
\SFSblue{Regarding MCMC estimation, the sparse finite mixture model is just a specific case of a finite mixture model, hence no modification of the MCMC scheme introduced in Section~\ref{sec:BayMix} is needed. With $\gamma_K\ll 1$, the sampler will automatically produce posterior draws with empty components, where $K_+< K$, while $K$ remains fixed.

For DPM models, where $K=\infty$, Steps~1 and 2 cannot be applied
immediately.  Even if conditional sampling is also possible for
infinite mixtures (e.g., with the slice sampler; \citealt{clips:Kalli2011}), the CRP sampler
is the most commonly used method. It produces posterior draws of the
number of clusters $K_+$ and the partition of the data and is
marginalized over the model parameters. The component-specific
parameters are then drawn conditional on the partition.}

\subsection{Point process representation of overfitting mixture models}
\label{sec:over}

For the case where the number of clusters is unknown, it is of
particular interest to assess how a mixture model with $K_+$
non-empty components 
is represented by an overfitting 
mixture model with $K > K_+$ components in the \PPR\ representation.

Let us consider, for example, that $K$ has one additional component,
i.e., $K = K_+ + 1$ implying that there is one redundant
component. The redundant component can either be included in the
mixture model by having a component with zero weight and arbitrary
component-specific parameters or by duplicating one component and
ensuring that the sum of the weights of the duplicated components
correspond to the true weight.

The \PPR\  of the true finite mixture distribution is invariant to adding
an \emph{empty} component, because the corresponding mark is zero and the
point is not visible.  On the other hand, if one component is split
into two components, then the position of the points remains
unchanged. In either case, the number of different points discernible
in the parameter space corresponds to the true number of non-empty components
$K_+$. In the overfitting case with duplicated components, one only
needs to ensure that the component weights are added for the identical
components.

The same behavior may be observed for the MCMC draws of an
overfitting mixture, \SFSblue{depending on how the hyper-parameters $\gamma_K$
for the Dirichlet prior 
$(\eta_1, \ldots, \eta_K) \sim {\cal D}_K(\gamma_K)$  of the component weights is
selected. 
On the one hand, if $\gamma_K$ 
is selected sufficiently small}, it is expected that the draws from filled components cluster around the true underlying parameter points whereas draws from empty components are draws from the prior which do not contain any data information and might be spread out over a wide area.
As explained in more
detail below, these draws can easily be identified by inspecting
their latent component assignments $(S_1, \ldots, S_N)$ and
removed. After their removal, one can focus on the component draws
from filled components and their identification. On the other hand, 
\SFSblue{if $\gamma_K$ 
is selected sufficiently large,}
an overfitting mixture with duplicated components is induced and the component distributions coincide for some components and need to be merged to
form a single cluster distribution. This indicates that an approach
pursuing overfitting with empty components provides a more
straightforward way to estimate the number of clusters (based on the
number of filled components) and identify the corresponding cluster
distributions (after removing component draws from empty components).

\subsection{The CliPS procedure when the number of clusters is unknown}\label{sec:CliPS-unknown-k_+}

As discussed in 
Section~\ref{sec:BayKunknon},
when the number of clusters is not known, the model specification
needs to be adapted to enable the estimation of the number of clusters
from the posterior of the number of filled components. In addition,
the post-processing of the MCMC draws needs to be adapted to account
for model specification 
uncertainty.
To this aim, the CliPS procedure 
introduced in Section~\ref{sec:CliPS-known-k_+}
for a mixture where the number of clusters is
known 
has to be adjusted.\\

\noindent \SFSblue{For a MFM model,
Step 1} is replaced with the following step:
\begin{enumerate}[Step 1*.]
\item 
Define a mixture model using a prior on $K$ and values
for the weight parameter $\gamma_K$ which enable to infer the number
of clusters from the posterior of the number of filled components.
	\end{enumerate}
	\SFSblue{For a sparse finite mixture model, 
Step 1 is replaced with the following step:
\begin{enumerate}[Step 1*.]
	\item 
	Define an overfitting finite mixture model with a large number  $K$ of components and choose a 
	value
	for the hyper-parameter $\gamma_K$ 
	of the weight distribution which encourages empty components \emph{a priori}.
\end{enumerate}
For both models}, an additional step has to be inserted between Step~2 and 3:
\begin{enumerate}[Step 2b.]
\item 
Post-process the MCMC draws:
\begin{enumerate}[(i)]
	\item Determine the posterior distribution of the number of filled
	components and estimate the number of clusters $\hat{K}_+$, e.g.,
	by taking the mode of the MCMC-based estimate of the posterior
	distribution $\Prob{K_+=k|\by}$.  This MCMC-based estimate is obtained
	by considering the number of non-empty components $ K_+^{(m)}$ for each
	iteration $m$, i.e., components to which observations have been
	assigned for this particular sweep of the sampler, 
	\begin{align} \label{drawKplus}
		K_+^{(m)} = K^{(m)} - \sum_{k=1}^{K^{(m)}} I[N_k^{(m)}=0],
	\end{align}
	where $N_k^{(m)}$  is the number of observations allocated to component
	$k$ and $I[]$ denotes the indicator function, and estimate
	the posterior $\Prob{K_+=k|\by}$ for each value $k = 1,\ldots, K$
	by the corresponding relative frequency.
	\item Eliminate all MCMC iterations where the number of filled
	components is 
	not equal to $\hat{K}_+$, i.e., the iterations
	$m \in \{1, \ldots, M\}$ where $K_+^{(m)}\ne \hat{K}_+$.
	\item For the remaining $M_{\hat{K}_+}$ MCMC iterations, exactly $\hat{K}_+$ components are filled. Eliminate
	the draws of all
	empty components such that only draws from
	$\hat{K}_+$ filled components remain for each of these MCMC
	iterations.
\end{enumerate}
\end{enumerate}
More specifically, for each of the $M_{\hat{K}_+}$ MCMC iterations with $\hat{K}_+$
filled components, no observations are allocated to the (redundant)
$K-\hat{K}_+$ components. Removing the draws from empty components in Step~2b(iii)
results in $M_{\hat{K}_+}$ draws from a mixture with exactly $\hat{K}_+$
non-empty components.  
Since this mixture can be treated as a mixture where the number of components is known, 
the further steps of the CliPS algorithm outlined in
Section~\ref{sec:CliPS-known-k_+} are applied to this sub-sample of 
$M_{\hat{K}_+}$
posterior 
draws to obtain a unique labeling.  Again the number $M_{{\hat{K}_+},\nu}$
of classification sequences not being a permutation is determined. All
$M_{\hat{K}_+}-M_{{\hat{K}_+},\nu}$ iterations where the classification sequence $\rho_m$
is a permutation of $\{1,\ldots,\hat{K}_+\}$ are considered for
further posterior inference.


\paragraph{An illustrative example}

We illustrate the CliPS procedure when the number of clusters is
unknown by revisiting Example~\ref{ex:known}, but this time assuming
that the number of clusters is not 
fixed \textit{a priori}. 
\begin{example}[\bf Unknown number of clusters]\label{ex:unknown}
We fit a dynamic Bayesian MFM model to the
artificial data set \SFSblue{previously}  used in
Example~\ref{ex:known}. Following
\citet{clips:Fruehwirth-Schnatter+Malsiner-Walli+Gruen:2021}, we use
the prior on the number of components given in \eqref{eq:BNB}
and the Dirichlet prior on the weights given in \eqref{eq:weights} with
$\gamma = 0.5$.  Otherwise the same prior specifications as in
Example~\ref{ex:known} are used. MCMC estimation is carried out via
the telescoping sampler using again the \textsf{R} package
\textbf{telescope}. 1,000 MCMC draws are recorded after omitting the
first 1,000 draws as burn-in.  We initialize the MCMC sampler in the
same way as for the case with known number of clusters.

The pairwise \PPR\ using all 15 combinations of mean
parameters in the six dimensions is visualized in
Figure~\ref{plot:example-fig3} (left) for all 1,000 MCMC
draws. The clustering of some component draws around the true
values is observed as well as the rather diffuse component draws
corresponding to draws from the prior distribution for empty
components.  Calculating the number of filled components for each of
the 1,000 draws indicates that all draws had 4 filled components
which thus corresponds to the posterior mode. Setting
$\hat{K}_+ = 4$ and only retaining MCMC draws with 4 filled
components implies that all 1,000 draws are retained.
After removing the draws of all empty components, the
pairwise \PPR\ of the means of the filled components only is
shown in Figure~\ref{plot:example-fig3} (right). 
\SFSblue{Notably, this plot is very similar to
	Figure~\ref{plot:example-fig2} suggesting that model identification
	is similarly straightforward.  Indeed, 
	applying $k$-means clustering results  in a non-permutation rate $\nu$ of zero.
	Finally,  the draws in Figure~\ref{plot:example-fig3} 
	are
	colored by the labeling obtained using CliPS, indicating the
	identified mixture.}
	\end{example}
	
	\begin{figure}[t!]
\centering
\includegraphics[width=0.49\textwidth]{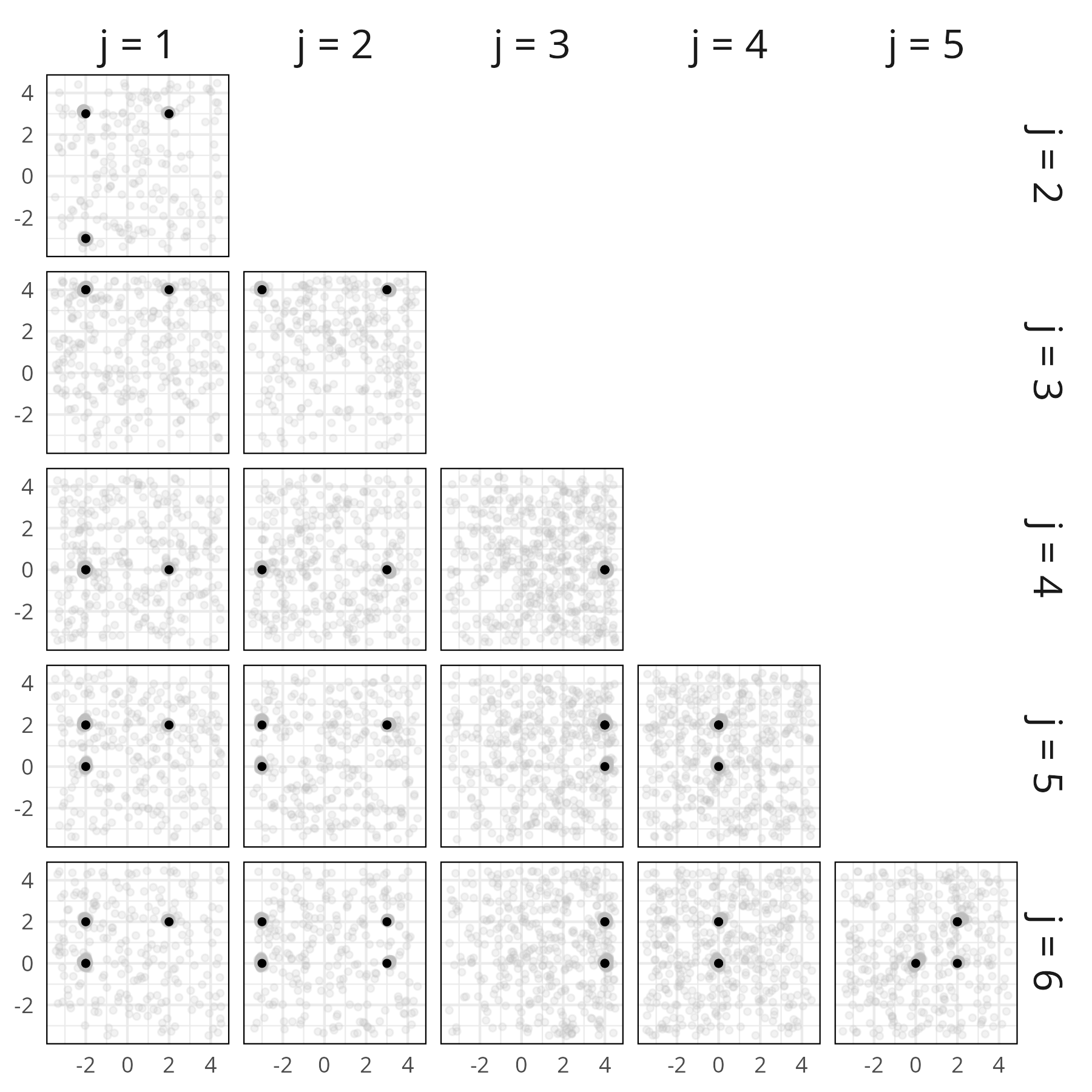}
\includegraphics[width=0.49\textwidth]{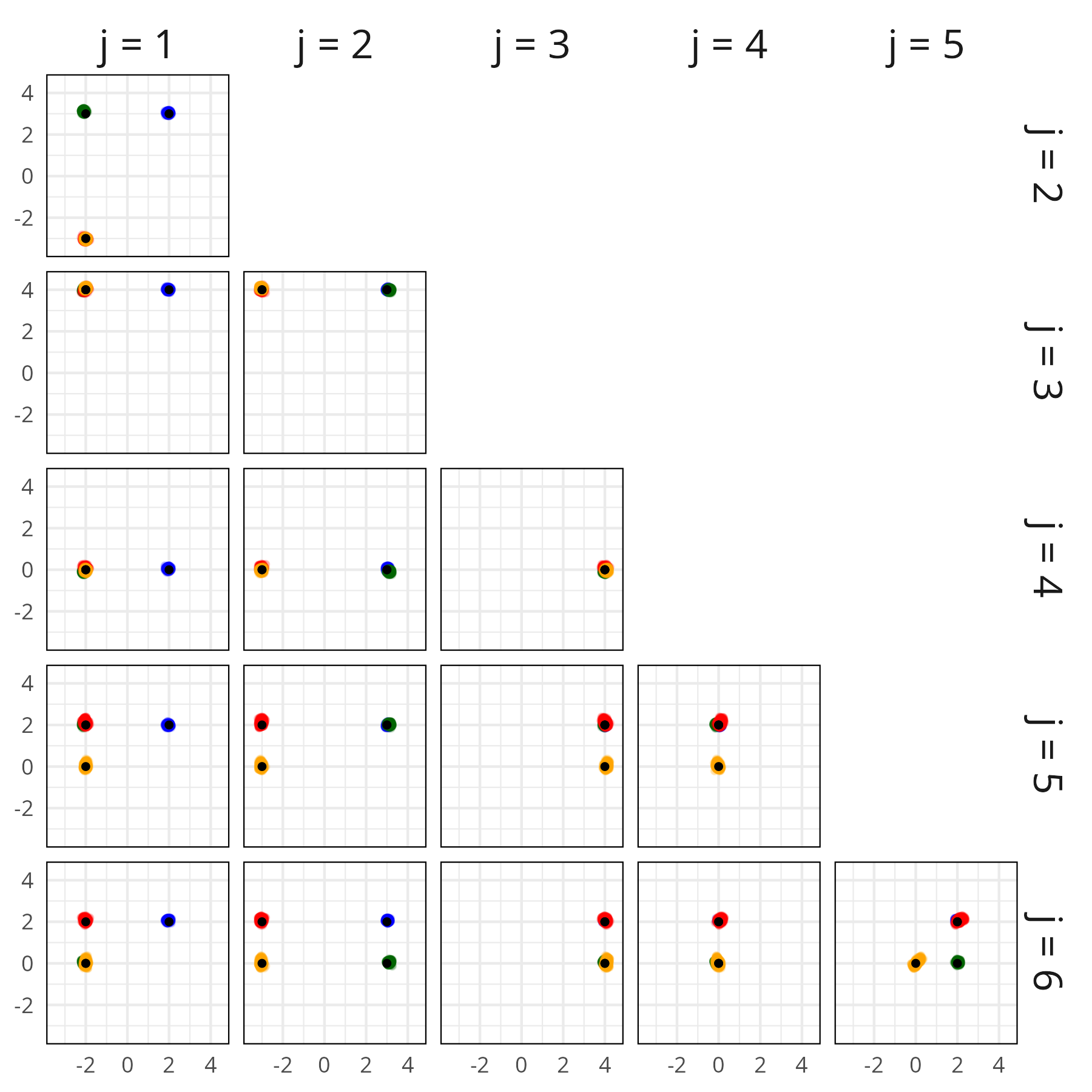}
\caption{{\bf Example 3} ($K$ unknown). Pairwise \PPR\ of the MCMC
	draws using the mean parameters in the various dimensions obtained
	for a Bayesian MFM model. Left: all MCMC draws including all
	components. Right: MCMC draws where $\hat{K}_+ = 4$ components are
	filled and retaining only filled components, with draws colored
	according to the labeling obtained with CliPS.}
\label{plot:example-fig3}
\end{figure}

\paragraph{\SFSblue{Implementing the CliPS procedure for a DPM model}}

\SFSblue{For a DPM model, 
Step 1 is replaced with the following step:
\begin{enumerate}[Step 1*.]
	\item 
	Define a DPM model with a hyper-prior on the concentration parameter $\gamma$. 
\end{enumerate}
Placing a prior on the concentration parameter ensures consistency in
estimating the number of clusters in a DPM model \citep{asc-etal:clu}
and allows the application of the CliPS procedure also in the context
of this specific infinite mixture.}

\SFSblue{MCMC estimation 
in Step~2 is entirely different 
for DPM models if the CRP sampler is applied, which is typically done even if a slice sampler could also be used. 
Since $K=\infty$, 
Step~2b  needs minor  modifications.
In particular, 
posterior draws of the number of filled components
$ K_+^{(m)}$  are directly available from CRP sampling and need not be recovered via \eqref{drawKplus}  as for finite mixtures. In addition, empty components are typically already removed from these draws.  
Apart from that, 
Step~2b remains the same, namely using in (i) an estimator $\hat{K}_+$ for the number of clusters based on the posterior distribution $\Prob{K_+=k|\by}$ and 
eliminating in (ii)  all MCMC draws, where the number of filled components differs from $\hat{K}_+$.
As for finite mixtures, 
the further steps of the CliPS algorithm outlined in
Section~\ref{sec:CliPS-known-k_+} are applied to this sub-sample of 
$M_{\hat{K}_+}$
posterior 
draws from a DPM model to obtain a unique labeling.}

\paragraph{\SFSblue{Using a different clustering algorithm}}

\SFSblue{In Step~3 of the CliPS procedure, a clustering algorithm is used to partition the
MCMC draws into $\hat{K}_+$ groups. We suggest to use $k$-means
clustering by default because usually this should be a rather
``easy'' clustering problem after having selected a suitable
functional. Other clustering algorithms could also be used, e.g.,
\citet{clips:Malsiner-Walli+Fruehwirth-Schnatter+Gruen:2016} use
$k$-centroid clustering based on the Mahalanobis distance to account
for non-spherical shapes of the parameter distributions.}

\SFSblue{Regarding its complexity, the $k$-means algorithm scales
linearly in the sample size, the number of groups and the dimension
of the input space.  For application in CliPS, the number of groups
will depend on the clustering application.  However, given that
$k$-means is also used for vector quantization, where many groups
are to be identified, the number of clusters is not expected to
represent a bottleneck. By contrast, the sample size and the
dimension of the input space may be influenced by not using all MCMC
draws to identify the group centers but thinning long MCMC chains
and carefully selecting a low-dimensional functional of the
component-specific parameters to characterize and differentiate the
clusters. Even if not all MCMC draws and all component-specific
parameters are used to identify the groups, all MCMC draws and
component-specific parameters may be relabeled based on the
$k$-means solution. Furthermore, many efficient implementations
including the selection of suitable initializations are available
\citep{Arthur+Vassilvitskii:2007, ikotun-et-al:2023}.}


\subsection{\SFSblue{Working with the CliPS procedure in practice}}

\SFSblue{Conceptually, the CliPS procedure relies on posterior separation of component-specific parameters
(or lower-dimensional functionals thereof) in the parameter space 
and the clustering procedure employed in Step~3 exploits  the strategy of  identifying a unique labeling of the  
component-specific parameters by  
permuting the component labels accordingly.}

\SFSblue{If $K_+$ is equal to the true number of clusters in a mixture with correctly specified cluster distributions,  the  component-specific parameters need to be distinct and  a unique labeling of the  
component-specific parameters  must exist theoretically. For such a mixture,   
using the entire parameter vector as functional in Step~3, i.e.,  $\funcTh{\bm{\theta}_k}=\bm{\theta}_k$,  will lead to a zero non-permutation rate  as the number $N$ of observations increases and the posterior distribution increasingly concentrates around the true parameter values.
However, if $K_+$ is overfitting the true number of clusters, then mathematically no unique labeling of the  
component-specific parameters exists, even for correctly specified cluster distributions. Label switching is intrinsic for such a mixture
and is revealed by the CliPS procedure by exhibiting  a high non-permutation rate, regardless of the choice of $\funcTh{\bm{\theta}_k}$.}  

\SFSblue{A key requirement for the CliPS procedure to work is that the number of clusters is estimated consistently. 
Consistency ensures that the posterior distribution of $K_+$  concentrates on the true number of clusters, justifying the use of the posterior mode as an estimator for $\hat{K}_+$.  As discussed by \citet{mil:con},  consistency holds in particular for MFM models, provided that the component densities are not misspecified.}

\SFSblue{Like many other procedures for identifying mixtures, the CliPS procedure faces challenges,  when  some clusters are quite similar and/or there is lack of heterogeneity across clusters in certain dimensions of the parameter space.  Depending on
the sample size $N$,  the component-specific posterior distributions might overlap in this case, resulting again 
in a non-negligible non-permutation rate $\nu$,
even if the true number of clusters has been estimated by the posterior mode. 
To filter out noise  injected by dimensions of the parameter space which do not carry relevant cluster-specific information,  we recommend to choose a lower-dimensional functional  $\funcTh{\bm{\theta}_k}$ of the component-specific parameter. This  functional  should be chosen based on domain-specific knowledge regarding which subset of parameters  is expected to distinguish the clusters. If the user's prior knowledge in this respect is limited,  care must be exercised not to remove parameter dimensions  that carry  relevant cluster-specific information. To reduce the dimensionality of $\funcTh{\bm{\theta}_k}$ also in this case, nuisance parameters such as the elements of the 
covariance matrix in a multivariate Gaussian mixture can be excluded from
$\funcTh{\bm{\theta}_k}$ nevertheless.} 

\SFSblue{As mentioned above, choosing a value for $\hat{K}_+$  which overfits the true number of clusters 
in  general results in a non-negligible non-permutation rate. This
happens in particular, if the component densities are misspecified. In this case, the  posterior distribution of $K_+$ will 
compensate for the lack of fit and 
the posterior mode is no longer 
a consistent estimator for $\hat{K}_+$.
Misspecification of the component densities typically leads to a vaguely 
dispersed posterior on $K_+$ and  
tends to blur the posterior separation of component-specific parameters,
yielding again a high non-permutation rate.  
If the mixture model was used for density estimation, one could account for  this model uncertainty by using Bayesian model averaging over all posterior draws, since  the estimated density is, by definition, insensitive to label switching and robust to non-zero non-permutations rates in the posterior draws for any specific
$K_+$.}

\SFSblue{Model-based clustering, on the other hand, relies on Bayesian model selection in terms of choosing a sensible  number of clusters which allows  identification of a unique labeling. 
If the posterior of $K_+$ lacks a clear mode, we recommend stratification of the MCMC sample by  all number of filled components  supported by the data 
which are then inspected to assess, whether one of them provides a suitable cluster
solution with a low non-permutation rate. 
As with any Bayesian model selection procedure, this naturally  means ignoring
competing solutions and might lead to a considerable reduction in the number of posterior draws which can be used for further inference. To resolve at least the second problem, we recommend to either rerun the procedure with a larger
sample size $M$ or to work instead with the CliPS procedure of Section~\ref{sec:CliPS-known-k_+}, where $K$ is equal to the selected number of clusters.  }

\SFSblue{Overall, in applied mixture analysis, 
the CliPS procedure 
reflects the suitability of the chosen model for clustering the data at hand.   Inspecting
the non-permutation rate induced by clustering in the \PPR\ of the MCMC draws highlights the
suitability of the fitted model for clustering and uniquely
characterizing the cluster distributions.  In case the fitted model
provides clear results with a very low non-permutation rate, this
suggests that the model induces a suitable cluster solution.}

\SFSblue{In case neither exploring different functionals nor stratifying the posterior draws yields  a cluster solution with low non-permutation
rate, this suggests that the cluster distributions are 
overlapping too heavily  to 
allow a unique
characterization of the clusters.
In such a situation, one 
should question the usefulness
of the cluster solution and 
may conclude that there is no
suitable cluster structure present in the data at hand. Alternatively, 
one might consider changing the model specification regarding
the prior and/or the component distribution of the mixture 
model to obtain a more
concentrated posterior distribution around a suitable cluster
solution.  The prior, e.g., can force sparser solutions through the choice of the hyper-parameter $\gamma_K$ (or its hyper-prior), see also the discussion in \citet{gre-etal:spy}. 
Flexible component distributions based  on a mixture-of-mixtures approach  considerably
increase robustness of the CliPS procedure  to misspecification of the component density, 
as demonstrated 
in \citet{mal-etal:ide, 
	mal-etal:wit}.}
	

	
	\section{Case studies}\label{sec:case-studies}
	
	We illustrate the application of the CliPS approach to a broad range of mixtures. The Bayesian mixture model specifications vary in regard to
	the component distributions, i.e., the clustering kernels, and the prior on $K$. In addition, different functionals are used for clustering in the \PPR .
	Specifically, we show inference using CliPS for a dynamic MFM of multivariate Gaussian distributions (with a discrete prior on $K$),
	for latent class analysis using a DPM (where $K$ is assumed to be infinite) and for a
	mixture of Markov chain models using a sparse finite mixture approach (with $K$ large, but fixed).
	
	\subsection{Dynamic mixture of finite mixtures modeling of
multivariate Gaussian distributions}\label{sec:dynam-mixt-finite}

In the following, we use the \texttt{diabetes} data set which contains
three metric variables corresponding to the measurements of glucose,
insulin, and sspg (steady state plasma glucose) and where a
classification into three groups of diabetes (chemical, normal, overt)
is known. The data set is available in the \textsf{R}
package \textbf{mclust}
\citep{clips:Scrucca+Fraley+Murphy:2023} and provides observations for
145 non-obese adult patients. The assumption is that the three metric
variables characterize the classes and hence the classes can be
inferred from the metric variables based on cluster analysis.  The
data set is visualized in Figure~\ref{plot:diabetes} (left) using
pairwise scatter plots. Even though the class distributions clearly
differ, there is also substantial overlap. Furthermore, the class
distributions seem to slightly deviate from multivariate Gaussian
distributions. Nevertheless, it still seems a viable assumption that a
model-based clustering approach using a multivariate Gaussian
distribution as clustering kernel is able to infer a cluster structure
which is well aligned with the known classification.

\begin{figure}[t!]
\centering
\includegraphics[width=0.48\textwidth]{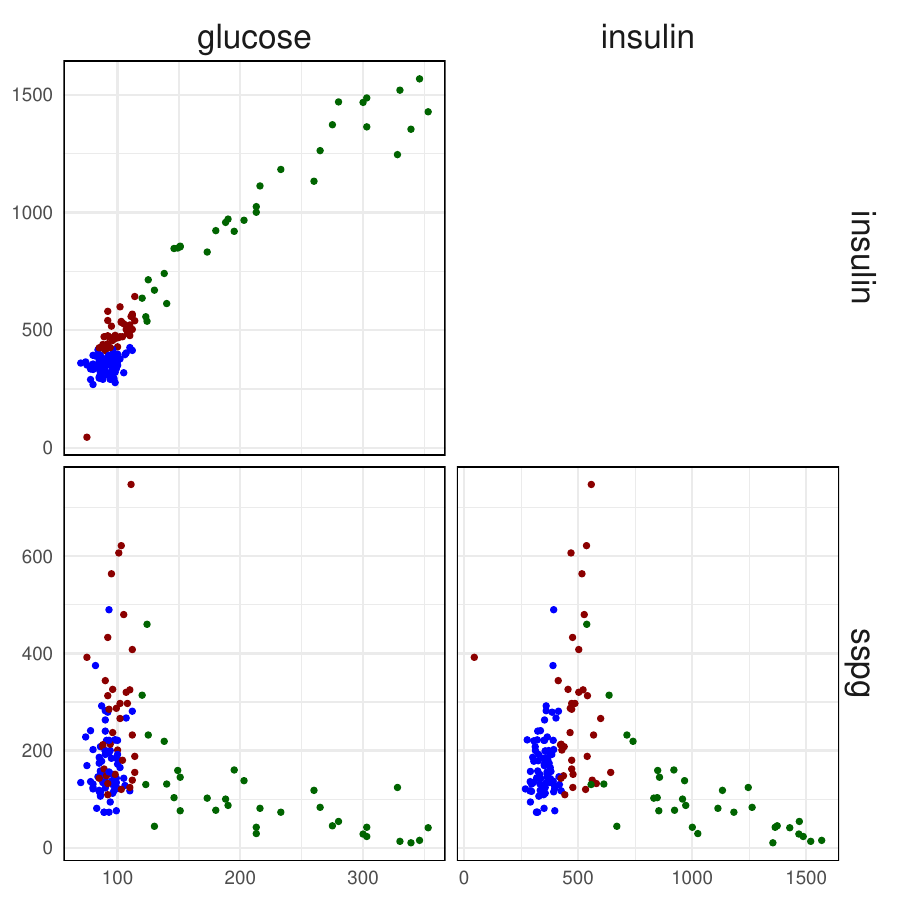}
\includegraphics[width=0.48\textwidth]{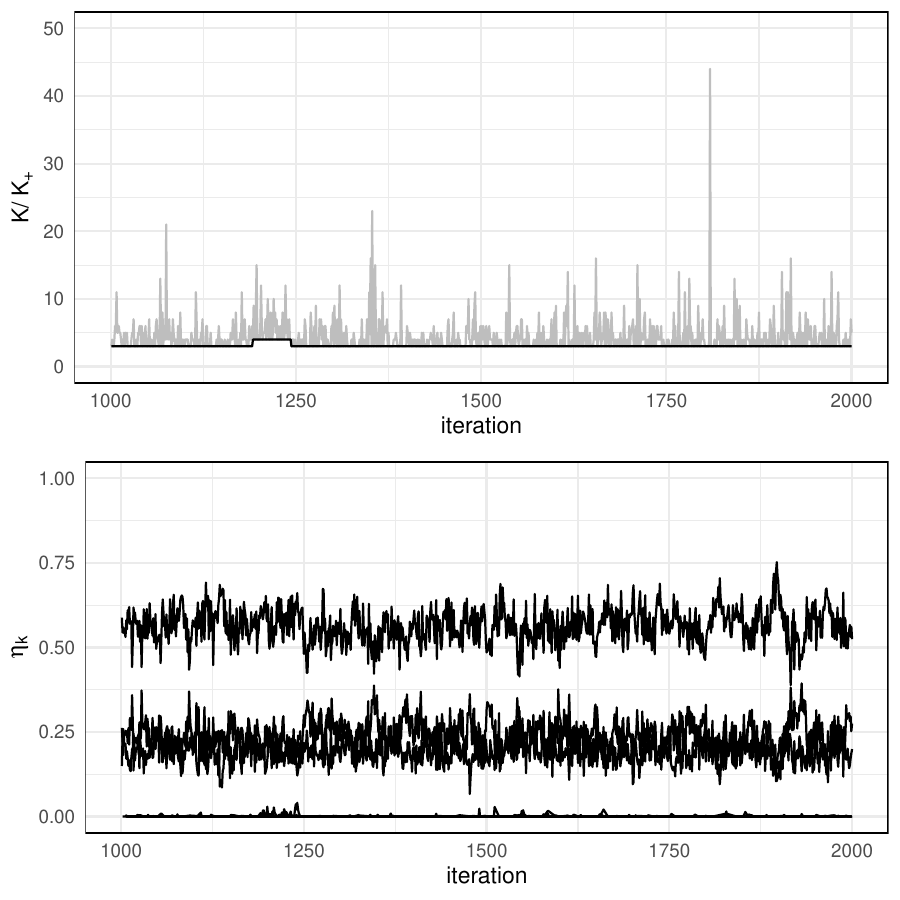}
\caption{Diabetes data. Left: pairwise scatter plots of the data
	colored by known classification. Right, top: trace plot of $K$
	(gray) and $K_+$ (black) for the recorded iterations. Right,
	bottom: trace plot of $\eta_k$ for $k =1, \ldots, K$.}
\label{plot:diabetes}
\end{figure}

We fit a dynamic MFM model with Gaussian
components $f_{\cN}(\cdot|\bmu_k,\bSigma_k)$ to perform unsupervised cluster
analysis, i.e., estimate the number of clusters and the cluster
distributions, and assess congruence between inferred clusters and the
known classification.  This means that we fit the same generative
model \SFSblue{as in Example~\ref{ex:unknown},
including all prior choices.} 
We record 1,000
MCMC iterations after dropping 1,000 iterations as burn-in when
using the telescoping sampler. The same initialization scheme as for
Example~\ref{ex:unknown} is used.

The trace plots of the sampled $K$ (in gray) and  the number $K_+$ of non-empty components among each  $K$ 
(in black) are shown in Figure~\ref{plot:diabetes} (right, top). 
In addition, Figure~\ref{plot:diabetes}
(right, bottom) provides trace plots of the sampled component
weights $\eta_k$ for $k=1, \ldots, K$. 
The
sampled $K$ values vary considerably and assume quite large values,
sometimes even greater than 20.
The induced $K_+$ values
are essentially always equal to three except for a short period where
four components are filled. This suggests that the mode of the
posterior of $K_+$ is clearly at $3$, and we thus estimate
$\hat{K}_+=3$ clusters. 


\begin{figure}[t!]
\centering
\includegraphics[width=0.48\textwidth]{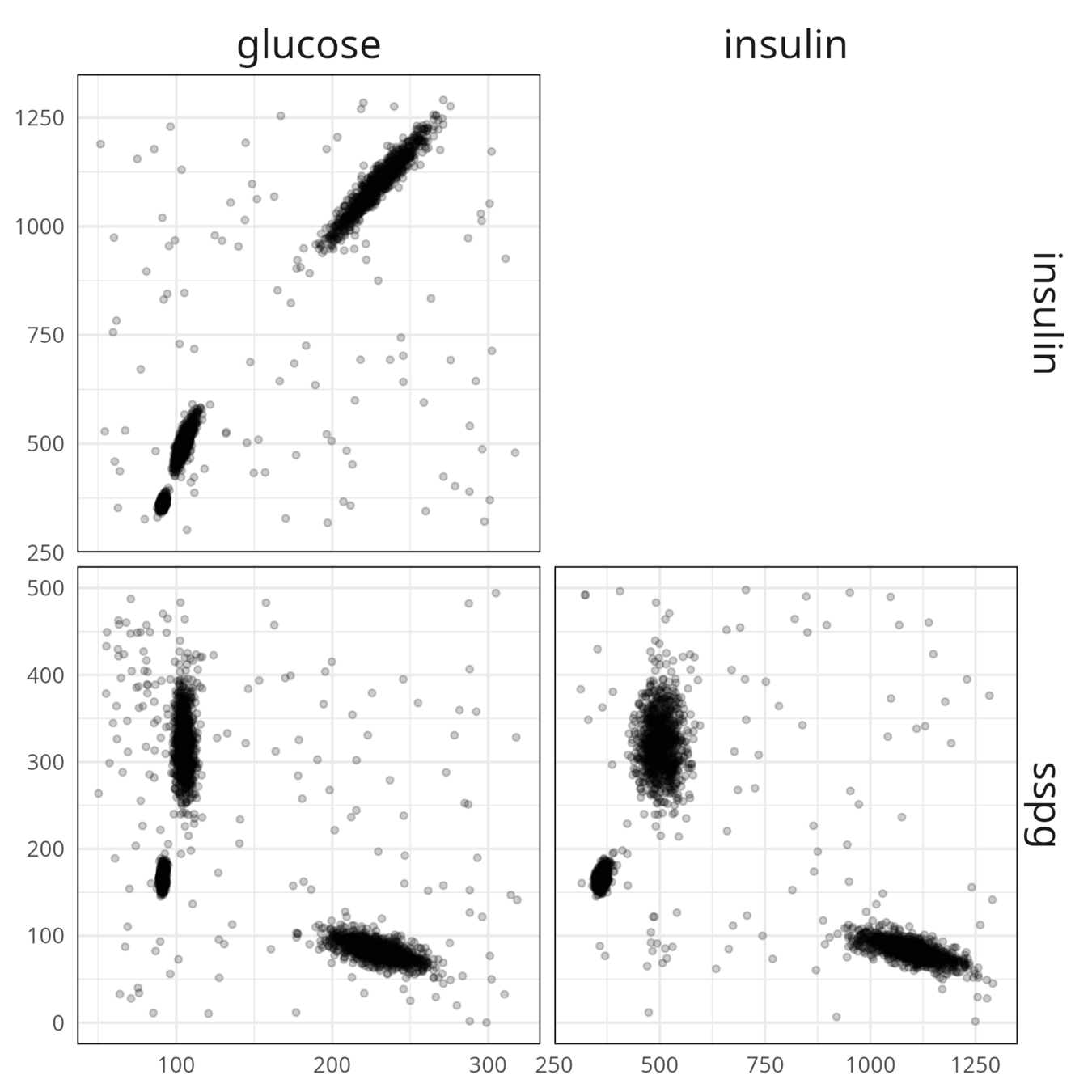}
\includegraphics[width=0.48\textwidth]{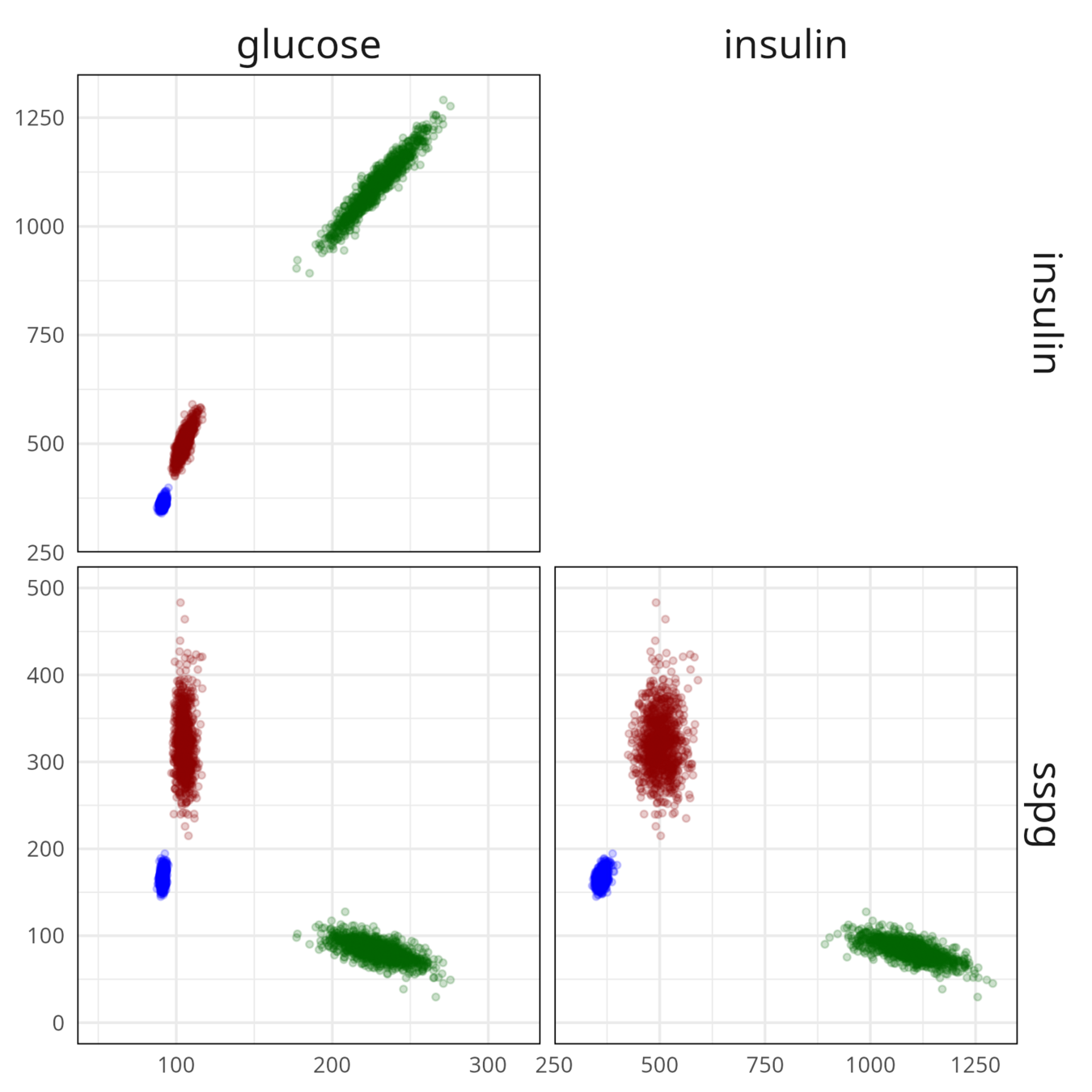}
\caption{Diabetes data. Left: pairwise \PPR\ of the mean
	parameters of all MCMC draws. Right: pairwise \PPR\ of
	the mean parameters of the MCMC draws where 3 components are
	filled with only draws of filled components retained and colored
	based on the labeling obtained with CliPS.}
\label{plot:diabetes-PPR}
\end{figure}


As before, we use the mean parameter vector of the Gaussian component distribution 
as
functional $\funcTh{\btheta_k}=\bmu_k$ of the 
component-specific parameter vector $\btheta_k=(\bmu_k, \bSigma_k)$
for the CliPS procedure. The 
pairwise \PPR\  of the component means
sampled from all components (also the empty ones) is shown in
Figure~\ref{plot:diabetes-PPR} (left). In these plots, draws from
``filled'' components which contain information about the cluster
distributions are overlaid by the misty cloud of draws sampled for
empty components  from the prior.

After retaining only the MCMC draws where the number of filled
components equals $\hat{K}_+ = 3$ and excluding from them the
component draws corresponding to empty components, the remaining
component draws of these MCMC draws are used to obtain the pairwise \PPR\  of the
mean parameters in Figure~\ref{plot:diabetes-PPR} (right). Clearly
three well-separated groups are discernible 
in every pairwise plot. The colors indicate the
solution obtained when clustering these component draws into
$\hat{K}_+=3$ groups using $k$-means clustering.  In this case, the
non-permutation rate $\nu$ is zero indicating that the posterior means
of the three components are well-separated.

Based on the classifications of $k$-means clustering, all draws are
relabeled and component-specific inference is performed. 
The posterior distributions of the mean parameters for each of the
clusters after identification can be used to characterize the
clusters.  Figure~\ref{plot:diabetes-PPR} (right) clearly indicates
that the green cluster differs most from the other two clusters with
respect to the mean values with the by far highest values for glucose
and insulin and the lowest values for sspg. The blue cluster has the
lowest mean values for glucose and insulin and also exhibits the
lowest posterior uncertainty.

Finally, observations may be classified based on the component they
have been assigned to most often during MCMC sampling. This results in
a final partition of the observations which can be compared to the
true class labels. This confusion matrix based on suitably relabeled
results has an accuracy of 0.855. \SFSblue{In addition, the adjusted
Rand index, which measures the agreement between two partitions
exceeding the expected agreement by chance given the marginals
scaled to take a value of 1 for two identical partitions
\citep{Hubert+Arabie:1985},} is 0.653.

\begin{figure}[t!]
\centering
\includegraphics[width=0.48\textwidth]{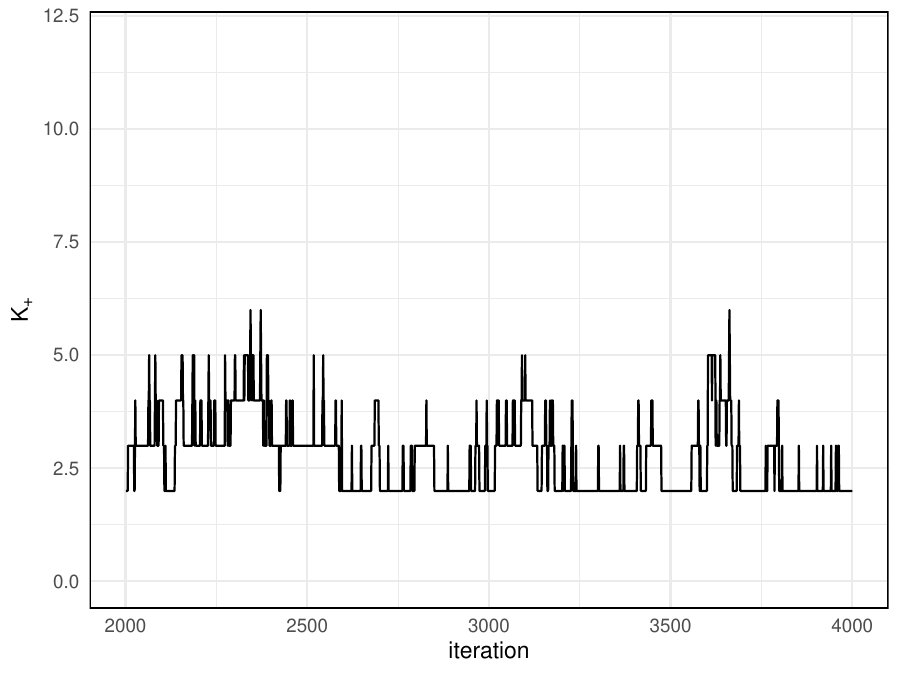}
\includegraphics[width=0.48\textwidth]{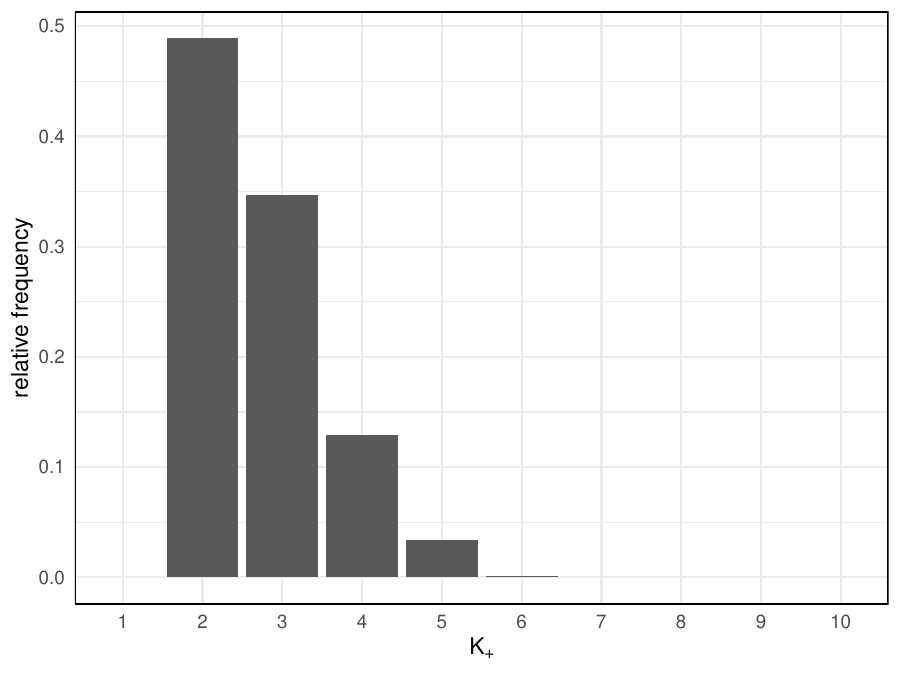}
\caption{Fear data. Trace plot (left-hand side) and
	posterior distribution (right-hand side) of $K_+$. {Note that $K$ is not plotted as it is assumed to be infinite.}}
\label{plot:fearinfiniteK}
\end{figure}

\subsection{Latent class model with a Dirichlet process prior}

The fear data set by \cite{clips:Stern1994} contains $N = 93$
observations of children who where tested in the context of infant
temperamental research. For each child, three categorical variables
(i.e., multivariate data of dimension $r = 3$) are observed, namely
fear of unfamiliar situations (fear) with $D_1 = 3$ categories,
fret/cry behavior (cry) with $D_2 = 3$ categories, and motor activity
(motoric) with $D_3 = 4$ categories.  The categories of the three
variables are 
ordered from weak to strong behavior. The
scientific hypothesis is that the behavior of the children can be
classified into two different profiles in children, which are referred
to as inhibited and uninhibited to the unfamiliar (i.e., avoidance or
approach to unfamiliar children, situations and objects).

The data are multivariate categorical data
$\ym_1, \ldots,\ym_N$, where $\ym_i= (y_{i1},\ldots,y_{i r})$
is the realization of an $r$-dimensional categorical random variable
$\bm{Y}=(Y_1, \ldots,Y_{r})$, with each element $Y_j$ taking one value
out of $D_j$ categories $\{1,\ldots, D_j\}$. Model-based clustering of
such data is performed by assuming the following multivariate mixture
distribution which corresponds to a latent class analysis model:
\begin{eqnarray}
p(\ym_i|\bm{\eta}_K, \bm{\pi}_K)= \sum_{k=1}^K \eta_k \prod_{j=1}^{r}\prod_{l=1}^{D_j} \pi_{k,jl} ^{I[y_{ij}=l]}
, \label{mix:lcmcat}
\end{eqnarray}
where $\pi_{k,jl} = \Prob{Y_{j}=l| S_i 
=k}$ is the occurrence probability
of category $l$ for feature $j$ in cluster $k$. The occurrence probabilities are summarized in
$\bm{\pi}_K = (\pi_{k,jl})_{k,j,l}$. This model relies on the
conditional independence assumption, i.e., the variables are independent given cluster
membership.

For the unknown probability distributions
$\bpi_{k,j}=(\pi_{k,j1},\ldots,\pi_{k,jD_j} )$ of feature $j$ in
cluster $k$ the symmetric, non-informative Dirichlet prior
$\bpi_{k,j}\sim \mathcal{D}_{D_j}(1) $ is specified.  In the spirit of
\cite{clips:Fruehwirth2019}, \SFSblue{a  \lq\lq sparse\rq\rq\ DPM model 
is defined, 
where the 
concentration parameter $\gamma$ follows the prior $\gamma \sim \cG(1,20)$.  }
To sample from the
posterior distribution the slice sampler
is used with tuning parameter $\kappa =0.5$.
After discarding 2,000 draws of burn-in, 2,000 iterations are
recorded.


\begin{figure}[t!]
\centering
\includegraphics[width=0.32\textwidth]{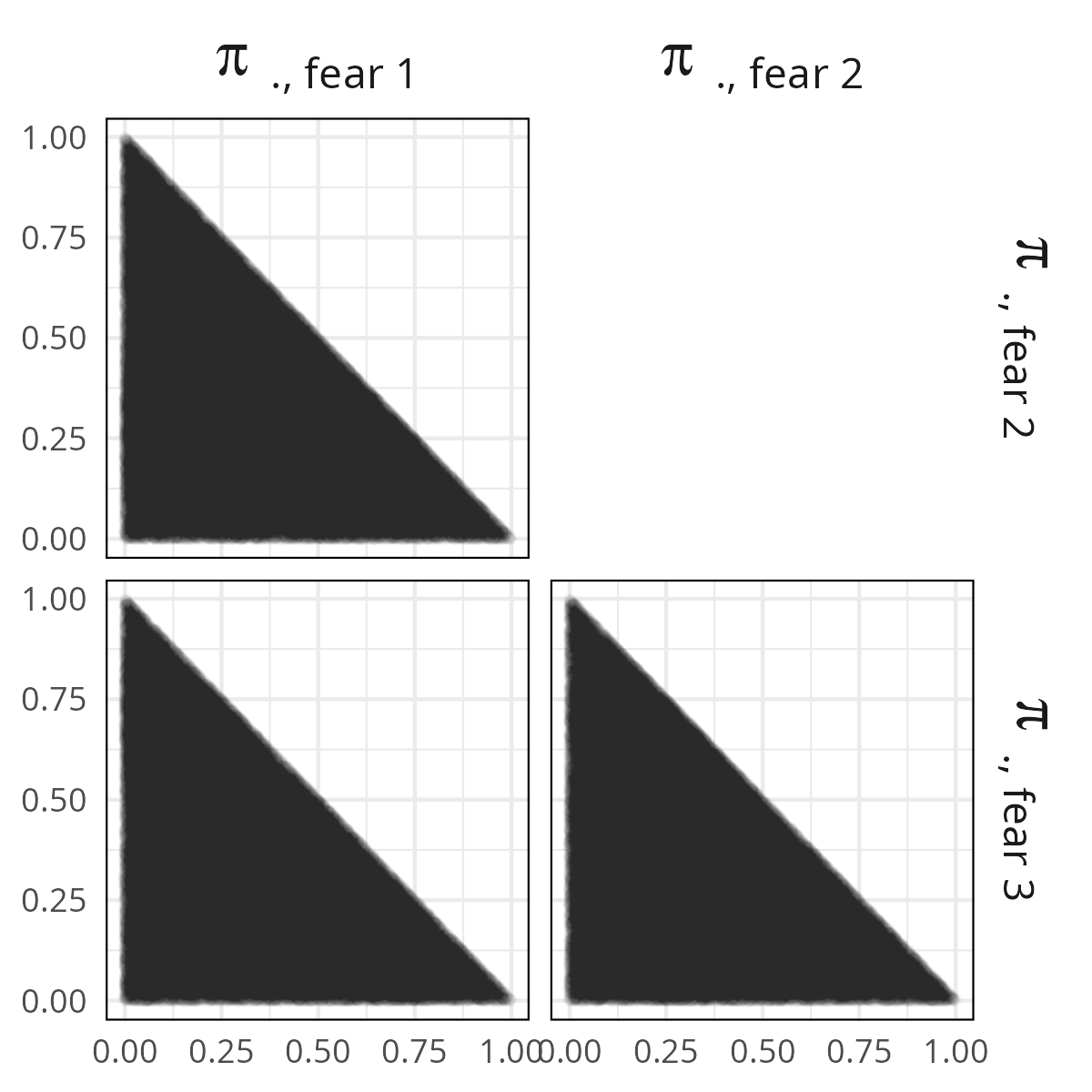}
\includegraphics[width=0.32\textwidth]{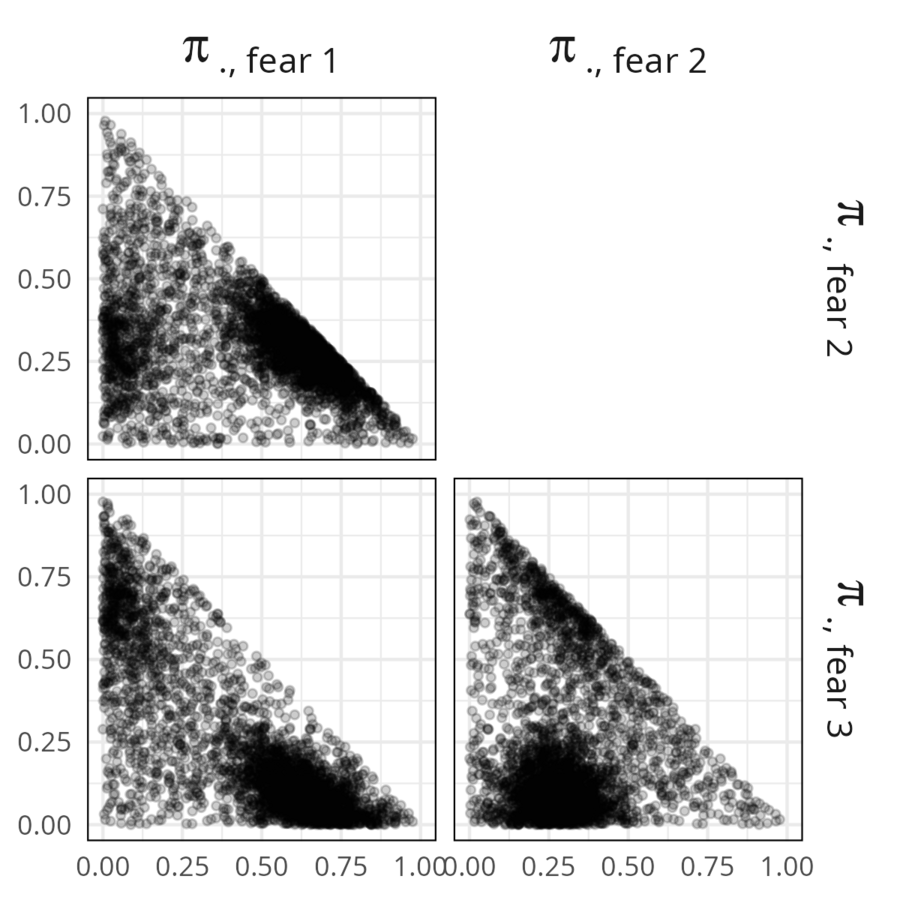}
\includegraphics[width=0.32\textwidth]{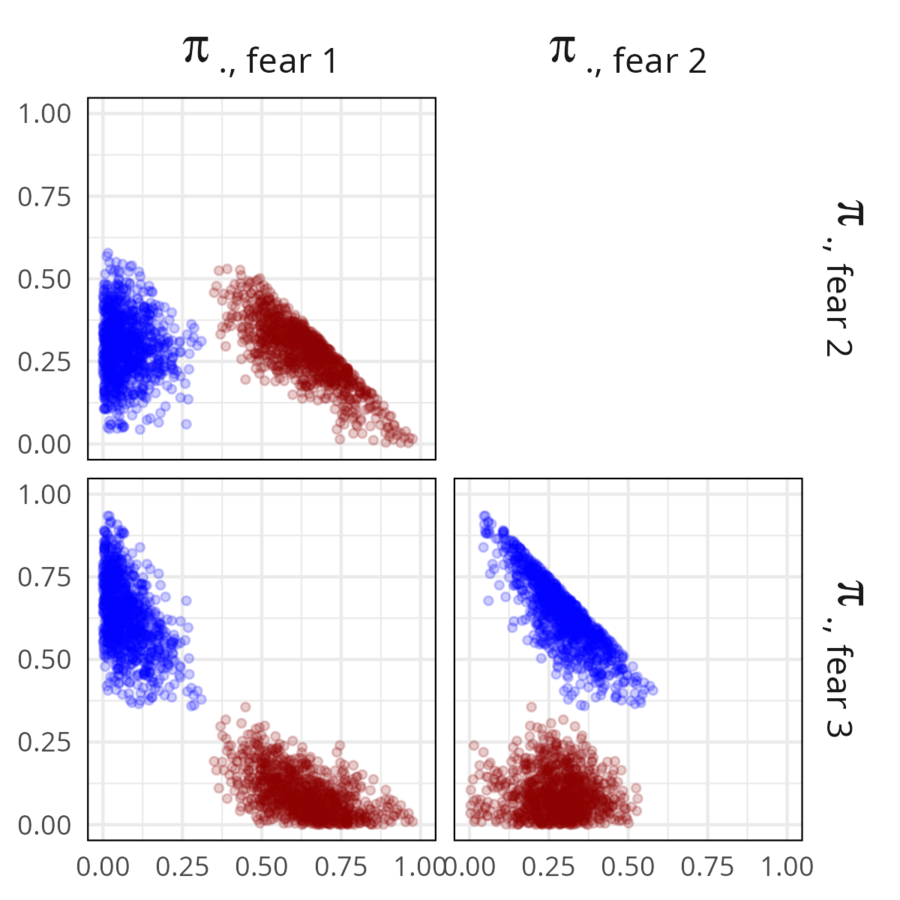}
\caption{Fear data. Pairwise 
	\PPR\ of the MCMC draws
	of the   occurrence probabilities 
	$\pi_{\cdot, \text{fear}1}$, $\pi_{\cdot, \text{fear}2}$ and $\pi_{\cdot,\text{fear}3}$ of the first variable
	(``fear'', 3 categories). Left: all draws including those from
	the empty components used in the slice sampler.
	\SFSblue{Middle: all draws from non-empty components.}
	Right: stratified
	draws for $K_+=2$ (without draws from empty components) colored
	based on the labeling obtained with CliPS.
} \label{plot:fearInfMix1}
\end{figure}


Figure~\ref{plot:fearinfiniteK} provides a trace plot of the posterior
draws of the number of filled components $K_+$ together with a bar
plot of the relative frequencies of the sampled $K_+$. These relative frequencies serve as an approximation of the
discrete posterior distribution of $K_+$. This
right-skewed posterior has its mode at $K_+=2$ with no support for
fewer, but considerable support for a larger number of
clusters. Using the mode as estimate for the number of clusters,
only the 978 MCMC draws with $K_+=2$ non-empty components are
considered in the subsequent analysis.

The entire ten-dimensional vector of all occurrence
probabilities $\btheta_k=$ $(\pi_{k,\text{fear}1}$,
$\pi_{k,\text{fear}2}$, $\pi_{k,\text{fear}3}$,
$\pi_{k,\text{cry}1}$, $\pi_{k,\text{cry}2}$, $\pi_{k,\text{cry}3}$,
$\pi_{k,\text{motoric}1}$, $\pi_{k,\text{motoric}2}$,
$\pi_{k,\text{motoric}3}$, $\pi_{k,\text{motoric}4})^\top$ is used
for clustering in the \PPR\ using the CliPS procedure, i.e.,
for this case study $\funcTh{\btheta_k}=\btheta_k$.  For
illustration, Figure~\ref{plot:fearInfMix1} (left) shows the pairwise
\PPR\ of the MCMC draws of the three occurrence probabilities
$\pi_{\cdot, \text{fear}1}$, $\pi_{\cdot, \text{fear}2}$ and
$\pi_{\cdot, \text{fear}3}$ of the first feature only, namely fear of
unfamiliar situations.  The plot shows the draws from \textit{all}
components, including the empty ones used in the slice sampler, for
all iterations $m=1, \ldots,M$, including those where
$K_+^{(m)}$ is different from two. No well-separated
parameter distributions can be discerned in this case. \SFSblue{After
omitting the draws from empty components, the parameters drawn still
cover the full simplex but two areas with higher density can be
distinguished in each scatter plot, see Figure~\ref{plot:fearInfMix1}
(middle).}  When retaining only the MCMC draws with
$K_+^{(m)}=2$ 
and omitting component draws from empty components, clearly two
well-separated clusters can be distinguished, see
Figure~\ref{plot:fearInfMix1} (right).

Figure~\ref{plot:fearInfMix1} shows that the support of the occurrence
probabilities for fear is the 3-dimensional simplex. In general,
parameter transformations are applied to such parameters in order to
map the parameters from the simplex to the unconstrained real space
before applying $k$-means clustering to improve the clustering
performance and respect the properties of the parameter space. However, in
the present case the cluster structure is so pronounced that even without
transformations $k$-means arrives at a suitable cluster solution. We
thus proceed without applying a parameter transformation, but suggest to
consider this in other applications.

\begin{table}[t!]
\centering
\begin{tabular}{rrrr|rrr|rrrr}
	\toprule
	\multicolumn{11}{c}{Estimated probabilities, after identification through CliPS}\\
	\midrule
	&\multicolumn{3}{c}{fear}&\multicolumn{3}{c}{cry}&\multicolumn{4}{c}{motoric}\\
	& 1 & 2 & 3 & 1 & 2 & 3 & 1 & 2 & 3 & 4 \\
	\midrule
	$k=1$ & 0.63 & 0.28 & 0.09 & 0.68 & 0.11 & 0.21 & 0.22 & 0.58 & 0.13 & 0.07\\
	$k=2$ & 0.07 & 0.29 & 0.63 & 0.27 & 0.30 & 0.43 & 0.15 & 0.17 & 0.40 & 0.28 \\ 
	\midrule
	\multicolumn{11}{c}{Benchmark probabilities by \citet{clips:Stern1994}}\\
	\midrule
	&\multicolumn{3}{c}{fear}&\multicolumn{3}{c}{cry}&\multicolumn{4}{c}{motoric}\\
	& 1 & 2 & 3 & 1 & 2 & 3 & 1 & 2 & 3 & 4 \\ 
	\midrule
	$k=1$ & 0.74 & 0.26 & 0.00 & 0.71 & 0.08 & 0.21 & 0.22 & 0.60 & 0.12 & 0.06 \\ 
	$k=2$ & 0.00 & 0.32 & 0.68 & 0.28 & 0.31 & 0.41 & 0.14 & 0.19 & 0.40 & 0.27 \\ 
	\bottomrule
\end{tabular}
\caption{Fear data. Occurrence probabilities of the benchmark by
	\citet{clips:Stern1994} and estimated using a latent class model
	with Dirichlet prior after model identification with
	CliPS.}\label{tab:fearInfMix}
	\end{table}
	
	When using $k$-means for clustering the MCMC draws $\btheta_k^{(m)}$ of all occurrence probabilities into two
	groups, the non-permutation rate $\nu$ is zero. This indicates that the
	estimated cluster distributions are easily distinguishable and
	well-separated. The coloring in
	Figure~\ref{plot:fearInfMix1} (right) illustrates the labeling
	obtained by the CliPS procedure for the three occurrence probabilities 
	of the first feature. 
	For the identified mixture model,
	Table~\ref{tab:fearInfMix} provides the estimated occurrence
	probabilities for all three features in the two identified clusters, together with the
	benchmark estimates reported by \citet{clips:Stern1994}. A clear
	congruence between the two solutions is visible.
	
	\subsection{Sparse mixture of Markov chain models}
	
	Finally, we apply CliPS to a sparse finite mixture of Markov chain models
	which is used to cluster categorical time series data
	\citep{clips:FruehwirthPamminger2010}. The data set
	\texttt{MCCExampleData} from the \textsf{R} package \textbf{bayesMCClust}
	\citep{clips:Pamminger:2012} contains data from the Austrian Social
	Security Database (ASSD), which combines detailed longitudinal
	information on employment and earnings by reporting wage categories
	(0--5, where 0 corresponds to no income) of $N=1,000$ male Austrian
	workers, who enter the labor market for the first time in the years
	1975 to 1985 and are less than 25 years old at entry. The cohort
	analysis is based on an observation period from 1975 to 2005. In
	Figure~\ref{plot:wage-data} (left), the sequence of income states over
	time is shown for four randomly selected individuals. The aim of the
	model-based cluster analysis is to identify groups of individuals with
	similar wage mobility behavior. {\Gblue The economic idea behind the analysis is to capture general patterns of wage mobility. Thus, interest lies in identifying few large groups of similar workers rather than many small groups of workers with specific patterns.}
	
	Let $y_{it},\; t = 1, \ldots, T_i$ be the sequence of categorical
	observations available for worker $i$ with the number of occasions
	$T_i$ varying over workers. Each $y_{it}$ takes $L$ potential states
	labeled by ${1,\ldots,L}$ and the individual time series
	$\ym_i = \{y_{i1},\ldots,y_{iT_i}\}$ are combined over workers
	$i=1,\ldots,N$.  This results in categorical time series data of
	different length for $N$ workers.
	
	Model-based clustering assumes that the data are obtained from $K_+$
	latent clusters which differ with respect to the cluster-specific
	parameter $\btheta_k$ of the clustering kernel $p(\ym_i|\btheta_k)$
	which is assumed to represent the data generation process for all time
	series in cluster $k$, $k = 1,\ldots,K_+$.  For clustering
	discrete-valued time series the first-order time-homogeneous Markov
	chain model is used as clustering kernel. It is characterized by the
	transition matrix $\bxi$, where
	$\xi_{jl} = \Prob{y_{it} = l|y_{i,t-1} = j}$, $j,l = 1,\ldots,L$. Each
	row of $\bxi$ represents a probability distribution over the discrete
	set $\{1,\ldots, L\}$, i.e., $\sum_{l=1}^L \xi_{jl} = 1$.  Hence, the
	cluster-specific parameter $\btheta_k$ is equal to $\bxi_k$ and the
	clustering kernel $p(\ym_i|\bxi_k)$ is given by:
	\begin{equation*}
p(\ym_i|\bxi_k) = \prod_{t=2}^{T_i} p(y_{it}|y_{i,t-1},\bxi_k) = \prod_{j=1}^L \prod_{l=1}^L \xi_{k,jl}^{N_{i,jl}},	
\end{equation*}
where $N_{i,jl}= \# \{y_{it}=l,y_{i,t-1} = j\}$ is the number of
transitions from state $j$ to state $l$ observed in time series $i$.

\begin{figure}[t!]
\centering
\includegraphics[width=0.32\textwidth]{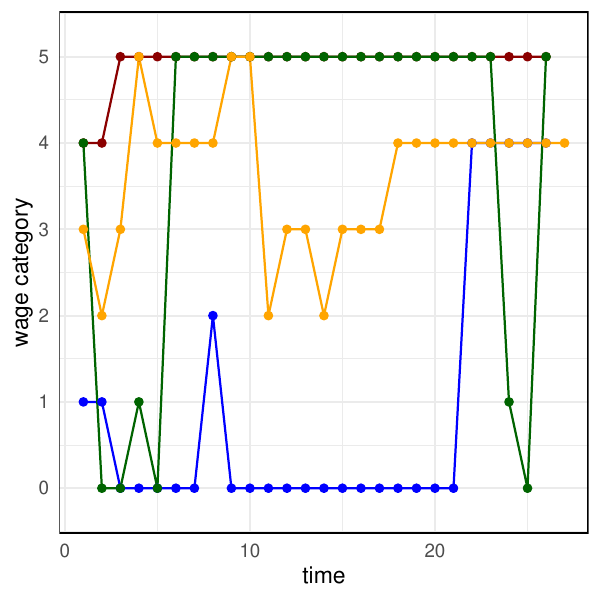}
\includegraphics[width=0.32\textwidth]{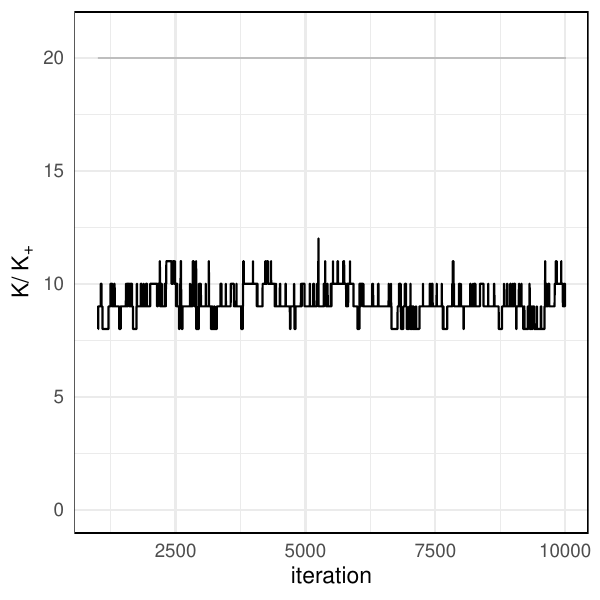}
\includegraphics[width=0.32\textwidth]{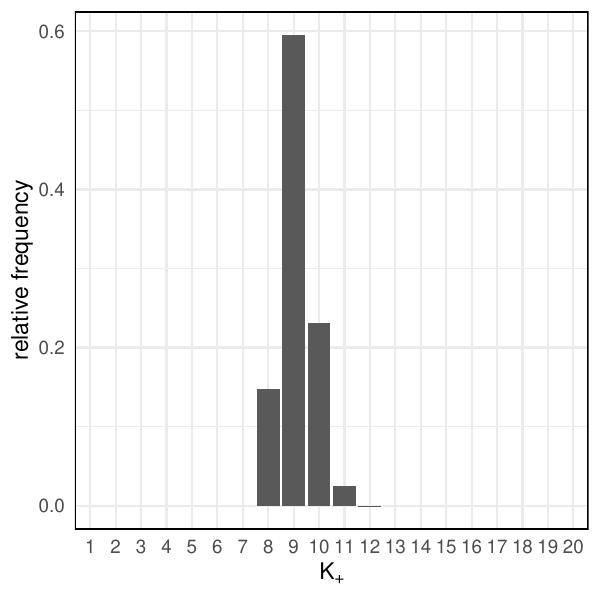}
\caption{Wage data. Left: time series of wage categories for four
	randomly selected individuals with time $t$ (in years) on the
	$x$-axis and wage category $l$ (with $l$ ranging from 0 to 5) on
	the $y$-axis.  Trace plot (middle) and posterior distribution (right) of
	$K_+$.} \label{plot:wage-data}
	\end{figure}
	
	We assume that the rows of $\bxi_k$ are \textit{a priori}  independent with each
	row $\bxi_{k,j}$ following an
	\SFSblue{$L$-dimensional Dirichlet
distribution, i.e.,
$\bxi_{k,j} \sim \mathcal{D} 
(\bm{\delta})$, with
$\bm{\delta}$ 
being an $L$-dimensional} vector of ones except for the entry corresponding to
the persistence probability which is set to zero.  We fit a sparse
finite mixture model
\citep{clips:Malsiner-Walli+Fruehwirth-Schnatter+Gruen:2016}  with the
number of components fixed at $K=20$ 
\SFSblue{to the data
using the package \textbf{bayesMCClust}. To encourage  close to zero weights in this overfitting mixture, we employ  the sparse symmetric
Dirichlet prior $\boldeta_K \sim \mathcal{D}_K(\gamma_K)$, where $\gamma_K=0.01$ implies strong prior concentration on the boundaries of the $K$-dimensional unit simplex.} 
We record $9,000$ iterations for
the MCMC sampler after discarding $1,000$ burn-in iterations.

Starting with an initial clustering with 20 filled components, rather
fast many of them become empty during MCMC sampling, and the number of
filled components switches between 8 and 11, see the trace plot in
Figure~\ref{plot:wage-data} (middle). The mode of the posterior is
9. However, for nearly half of these draws only a single observation
is assigned to one filled component and for about a third of these
draws two observations are assigned to one filled component. 
We do not use all elements of the component-specific transition matrix 
as a functional
for clustering in the \PPR\ using the CliPS procedure. Instead of clustering all 30 free elements
of $\bm{\xi}_k$, given that the row entries sum to 1, we use only the
six persistence probabilities, i.e., 
$\funcTh{\btheta_k}=(\xi_{k,11}, \ldots, \xi_{k,66})^\top$ is equal to 
the diagonal elements of
$\bm{\xi}_k$. The persistence probabilities are expected to differ
across clusters based on clusters varying in regard to wage mobility.

\SFSblue{For $\hat{K}_+=9$, 
the \PPR\ of all diagonal elements of the transition matrices showed  diffuse draw clouds with a 
non-permutation rate of 8\%. 
As interest lies in capturing few large clusters,
we decided to investigate  $\hat{K}_+=8$ clusters which still receives considerable posterior probability in 
Figure~\ref{plot:wage-data}. 
Figure~\ref{plot:wage-PPR-sub} displays in the upper triangle the
pairwise \PPR\ 
of the posterior draws of the persistence probabilities of
all MCMC draws with eight filled components and with all draws from empty
components excluded. Clearly several density clusters are
discernible in the scatter plots even though based on only these
bivariate projections 
no clear separation is visible. However, using
$k$-means on the full 6-dimensional space 
in the CliPS procedure results in 
a very small non-permutation
rate of 1.4\%. This implies that
$1,328$ posterior draws are retained for further inference. The labeling induced by $k$-means clustering is
shown in the lower triangle of Figure~\ref{plot:wage-PPR-sub}.


\begin{figure}[t!]
	\centering
	\includegraphics[width=\textwidth]{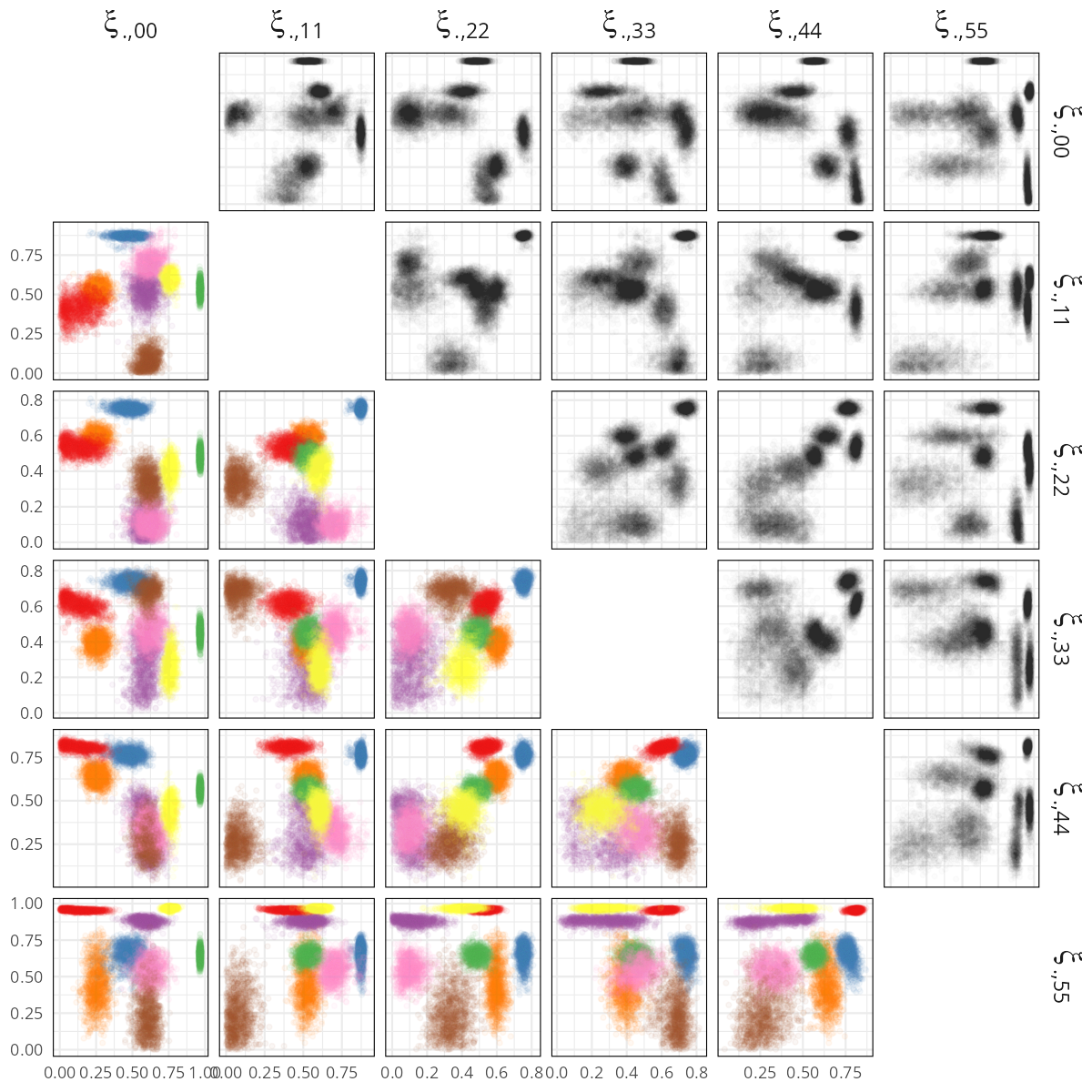}
	\caption{Wage data. Pairwise \PPR\ of the MCMC draws of the
		persistence probabilities where eight components are filled, only
		draws from filled components are kept and only draws are retained
		where a unique labeling could be obtained.  The colors in the
		lower triangle reflect the labeling obtained with CliPS.
	} \label{plot:wage-PPR-sub}
\end{figure}


\begin{figure}[t!]
	\centering
	\includegraphics[width=\textwidth]{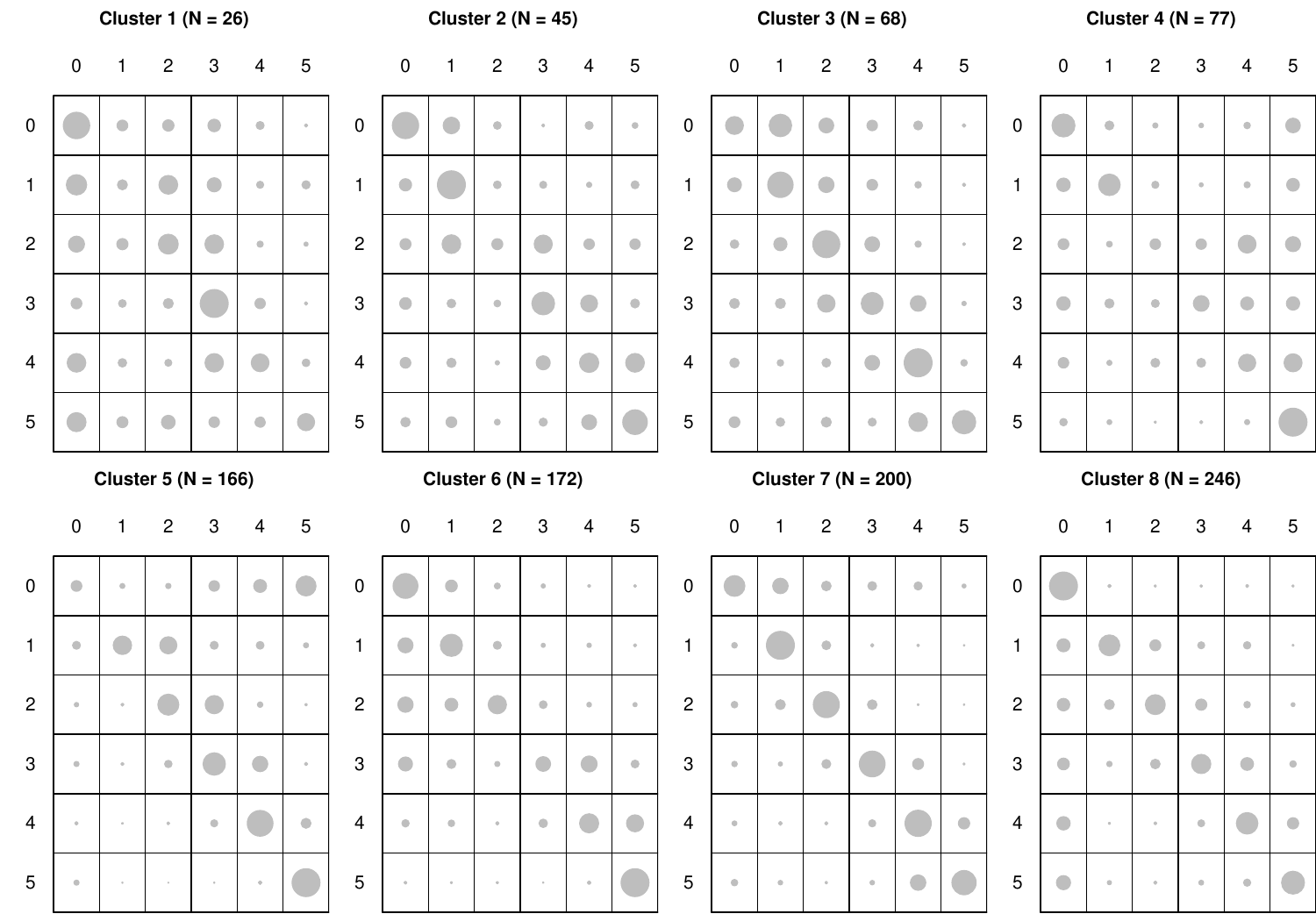}
	\caption{Wage data. Estimated transition probabilities.}\label{plot:wage-xi}
\end{figure}

When estimating  eight clusters, both the very low non-permutation rate as well as the well-separated posterior parameter distributions shown in Figure~\ref{plot:wage-PPR-sub} indicate that $\hat{K}_+=8$  
is an appropriate number of clusters for these data.}
Based on the model identification obtained with CliPS, the estimated
transition matrices for the eight cluster distributions are shown in
Figure~\ref{plot:wage-xi}. It can be seen, for
instance, that 
individuals in cluster 1 are likely to get unemployed, as they have a
high probability to move  to state 0 from all other states. In contrast,
persons in cluster 5 have a high probability to either keep their earnings, 
to move up to higher earnings in case they have a positive income
(i.e., categories 1--5) or to  move from the zero class to a
higher income class. Figure~\ref{plot:wage-xi} also contains the
cluster sizes obtained from the classification of the observations
when assigning them to the component they have been assigned to most
often using the identified MCMC samples. 

Figure~\ref{plot:wage-typical} shows the seven most typical
individuals of each cluster, selected based on their \textit{a posteriori}
probabilities after model identification with CliPS. These profiles
obviously are quite different between the clusters.

\begin{figure}[t!]
\centering
\includegraphics[width=\textwidth]{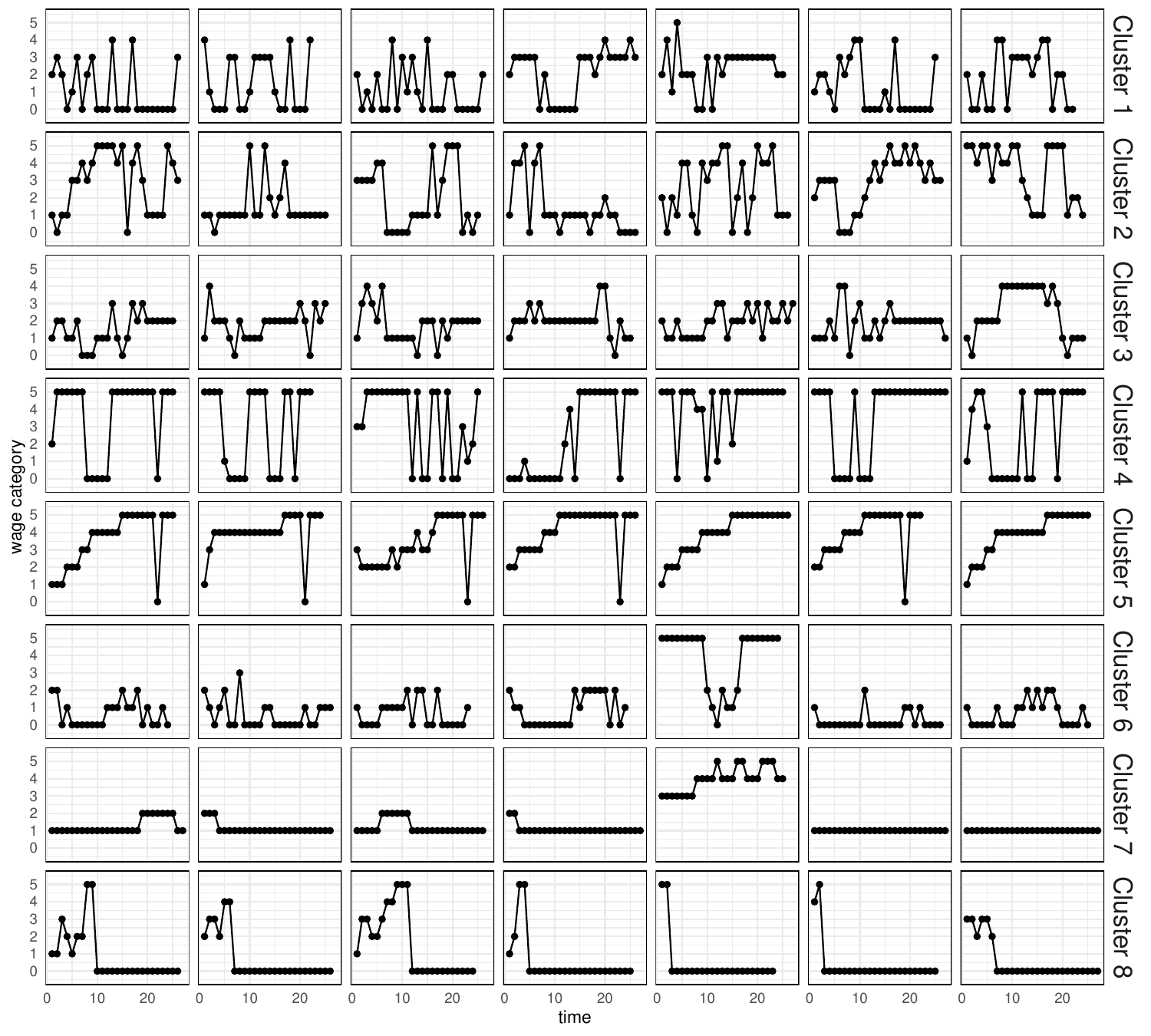}
\caption{Wage data. The eight rows show the 
	seven most typical members 
	for each of the eight clusters identified by the CliPS procedure.
} \label{plot:wage-typical}
\end{figure}


\section{Discussion and conclusion}\label{sec:disc-concl}

Model-based clustering aims at identifying cluster distributions as
well as assigning observations to clusters. We reviewed the use of
Bayesian 
mixture models and a sampling approach based on data augmentation, i.e.,
including the unknown cluster memberships as latent variables, for
fitting the model to data in this context.
Based on the MCMC draws of a mixture model, the CliPS procedure obtains an
identified model by clustering a functional of the component-specific
parameter draws using, for example, $k$-means clustering. This approach has several attractive features.

The CliPS procedure is applicable to any kind of clustering kernel. 
In the present paper, we considered
multivariate Gaussian mixtures, latent class analysis, and mixtures
of Markov chains.
The core idea of the procedure, namely to perform clustering in the
parameter space of (a functional of) the component-specific
parameters to identify a unique labeling, 
is very generic and 
has been 
applied previously
to a wide range of alternative 
mixture models, including 
mixtures of random effects models
\citep{fru-etal:bay}, uni- and multivariate skew-N and skew-$t$
mixtures 
\citep{fru-pyn:bay}, mixtures of Gaussian mixtures
\citep{mal-etal:ide}, and, very recently,  mixtures of latent class models  \citep{mal-etal:wit} and 
mixtures of factor analyzers
\citep{gru-fru:dyn}. 
Applications in a time series context include  switching state space models
\citep{fru:ful} and 
model-based clustering of  time series \citep{fru-kau:mod,vav-etal:mod}.

In the present paper, we outlined in detail the CliPS procedure in
the context of Bayesian model-based clustering when the number of
clusters is known as well as when it is unknown.  In case the number
of clusters is unknown, a suitable Bayesian mixture model needs to
be specified which allows to infer the number of clusters from the
number of filled components during MCMC sampling.  This was
illustrated for a MFM model, a DPM model and a sparse finite mixture
approach, but it can be applied to even more general Bayesian
non-parametric mixtures such as Pitman-Yor process mixtures
\citep{mul:bay}.

Furthermore, the CliPS procedure can  be applied to common
generalizations of the standard finite mixture model discussed in
this paper, namely hidden Markov and Markov switching models
\citep{hah-etal:mar,kau:sta} as well as mixture of experts models
\citep{fru-etal:lab,gor-fru:mix}, provided that the number of
components or states $K$ is known. Again, the core idea --
to cluster (a functional of) the
component-specific parameters in the parameter space to identify a unique labeling -- is
exactly the same. Based on a suitable MCMC sampler for these models,
Step~3 to Step~5 of the CliPS procedure are basically identical, with the minor
modification in the last step that the parameters defining the prior
on the latent \SFSblue{assignment variables} $S_1, \ldots,S_N$, such as the transition matrix of
a hidden Markov model, have to be relabeled instead of the weight
distribution $\eta_1, \ldots, \eta_K$.

Finally, the CliPS procedure 
allows to reflect the suitability of the chosen model for clustering the data at hand.   Inspecting
the non-permutation rate induced by clustering in the \PPR\ of the MCMC draws highlights the
suitability of the fitted model for clustering and uniquely
characterizing the cluster distributions.  In case the fitted model
provides clear results with a very low non-permutation rate, this
suggests that the model induces a suitable cluster solution.

\section*{\SFSblue{Declarations}}

\subsection*{\SFSblue{Data Availability}}

\SFSblue{The data sets used in the three cases studies are available
from \textsf{R} packages hosted on the Comprehensive \textsf{R}
Archive Network (CRAN). The \texttt{diabetes} data set is contained
in package \textbf{mclust}; the \texttt{fear} data set is defined in
the vignette ``Bayesian Latent Class Analysis Models with the
Telescoping Sampler'' contained in package \textbf{telescope} and
the \texttt{MCCExampleData} data set is contained in
\textbf{bayesMCClust}.}

\subsection*{\SFSblue{Code Availability}}

\SFSblue{Package \textbf{telescope} contains a vignette entitled
``Using CliPS to Identify a Dynamic Mixture of Finite Mixtures of
Multivariate Gaussian Distributions -- Case Study: Diabetes Data''
which provides replication material for the case study using the
\texttt{diabetes} data set in Section~\ref{sec:dynam-mixt-finite}.}

\bibliographystyle{Chicago}
\bibliography{clips, bayesian-mixtures}
\end{document}